\documentstyle[12pt,epsf,eqsecnum,aps,amssymb,amsfonts]{revtex}

\newlength{\ul}
\newcommand{\Loop}{\!\!\!\raisebox{-0.4\ul}
{\setlength{\unitlength}{0.1\ul}
\begin{picture}(12,10)(0,0)
\thicklines
\put(1,5){
\put(5,0){\circle{10}
\put(0,-5){\circle*2}}
}\end{picture}}}
\newcommand{\Dashloop}{\!\!\!\raisebox{-0.4\ul}
{\setlength{\unitlength}{0.1\ul}
\begin{picture}(12,10)(0,0)
\thicklines
\put(1,5){
\put(5,0){\circle{10}
\put(0,-5){\circle*2}
\put(-5,0){\dashbox{0.7}(10,0){}}}
}\end{picture}}}
\newcommand{\Leftline}{\!\!\!\raisebox{-0.4\ul}
{\setlength{\unitlength}{0.1\ul}
\begin{picture}(12,10)(0,0)
\thicklines
\put(1,5){
\put(10,0){\circle*2}
\put(0,0){\line(1,0){10}}
}\end{picture}}}
\newcommand{\Rightline}{\!\!\!\raisebox{-0.4\ul}
{\setlength{\unitlength}{0.1\ul}
\begin{picture}(12,10)(0,0)
\thicklines
\put(1,5){
\put(1,0){\circle*2}
\put(8.5,-1.3){x}
\put(0,0){\line(1,0){10}}
}\end{picture}}}
\newcommand{\Dashleftline}{\!\!\!\raisebox{-0.4\ul}
{\setlength{\unitlength}{0.1\ul}
\begin{picture}(14,10)(0,0)
\thicklines
\put(1,5){
\put(12,0){\circle*2}
\put(4.5,0){\dashbox{0.7}(4,4){}}
\put(0,0){\line(1,0){12}}
}\end{picture}}}
\newcommand{\Dashrightline}{\!\!\!\raisebox{-0.4\ul}
{\setlength{\unitlength}{0.1\ul}
\begin{picture}(14,10)(0,0)
\thicklines
\put(1,5){
\put(1,0){\circle*2}
\put(10.5,-1.3){x}
\put(4.5,0){\dashbox{0.7}(4,4){}}
\put(0,0){\line(1,0){12}}
}\end{picture}}}

\begin{document}

\sloppy

\draft

\title{Correlation functions near Modulated and Rough Surfaces}
\author{\\Andreas Hanke$^1$ and Mehran Kardar$^{1,2}$}
\address{
$^1$Department of Physics, Massachusetts Institute of Technology,\\
Cambridge, Massachusetts 02139\\
$^2$Institute for Theoretical Physics, University of California,\\
Santa Barbara, California 93106}
\date{\today}
\maketitle

\bigskip

\medskip

\begin{abstract}
In a system with long-ranged correlations, the behavior of
correlation functions is sensitive to the presence of a boundary. 
We show that surface deformations strongly modify this behavior as 
compared to a flat surface. The modified near surface correlations 
can be measured by scattering probes. To determine these correlations, 
we develop a perturbative calculation in the deformations in height from 
a flat surface. Detailed results are given for a regularly patterned 
surface, as well as for a self-affinely rough surface with roughness 
exponent $\zeta$. 
By combining this perturbative calculation in height deformations
with the field-theoretic renormalization group approach, 
we also estimate the values of critical exponents governing the behavior 
of the decay of correlation functions near a self-affinely rough 
surface. We find that for the interacting theory, a large enough 
$\zeta$ can lead to novel surface critical 
behavior. We also provide scaling relations between roughness induced 
critical exponents for thermodynamic surface quantities.
\end{abstract}

\bigskip

\pacs{PACS numbers: 68.35.Rh, 68.35.Ct, 05.70.Jk, 64.60.Fr}

\narrowtext

\newpage

\section{Introduction}

In a material with long-ranged correlations, such as a liquid
crystal or a superfluid,
any local perturbation has influence over large distances. As a 
result, local properties, such as magnetization density, as well 
as correlation functions are modified on approaching a surface.
{\em Critical\/} behavior near surfaces or defects, 
which is quite different from the bulk, has been extensively studied 
by means of the field-theoretic renormalization-group approach 
\cite{Bin83,Die86,DW84,AH2000}. In this case, the local order
parameter $\Phi$ is perturbed near the surface up to a distance
set by the diverging bulk correlation length
$\xi \sim |T - T_c|^{- \nu}$, where $T_c$ is the bulk critical 
temperature. Theoretical predictions for surface criticality
have been tested experimentally \cite{MDPJ90,Dosch92,BPH93,exp,Law2001} 
and in simulations \cite{LB90,RDW92}.
In particular, the grazing incidence of x-rays and neutrons 
\cite{DW84} has become a standard tool for 
probing critical behavior near surfaces and interfaces 
\cite{MDPJ90,Dosch92,BPH93,exp}. For instance, the decay of the 
two-point correlation function has been measured close to the 
surface of a Fe$_3$Al crystal near its continuous order-disorder 
transition by the method of grazing incidence of x-rays \cite{MDPJ90}.
The phenomenon of critical adsorption near columnar defects \cite{AH2000} 
has apparently been observed by small angle scattering of light 
in a NH$_4$Br crystal near a continuous structural phase transition
\cite{BV80}.

Most theoretical investigations so far have been restricted to
{\em flat\/} surfaces. This is justified to a certain degree, since
microscopic deviations from this idealized picture such as terraces 
of monoatomic height do not change the universal surface critical 
behavior \cite{PS98,Diehl98}. However, for deviations on mesoscopic 
length scales, new phenomena are expected. Such deviations can be 
divided into two classes:

(i) Advanced experimental methods of nanoscience such 
as x-ray \cite{Tolfree}, guided growth \cite{Whi97}, and nanosphere
lithography \cite{BSK98}, allow one to endow surfaces with specific, 
regular geometrical patterns down to the nanometer scale. 
These structures hold much promise for applications 
towards nanochips \cite{LXQ96} or optoelectronic devices \cite{Fen96}. 
The surface modulations also offer a wide range of 
possible applications in fluid environments. 
For instance, at temperatures between the wetting 
temperature $T_w$ of the corresponding planar substrate and 
the critical temperature $T_c$ of the bulk fluid, one can manipulate
the adsorption properties of the fluid on the substrate by endowing
the surface with periodic patterns of various shapes 
\cite{Die99,RP2000}.

(ii) Surfaces or interfaces can be naturally rough, e.g., 
due to growth, fracture, or erosion. 
One possibility is that the substrate has a {\em fractal\/} surface, 
so that the surface area $S$ grows as a power of the projected area, 
i.e., $S \sim L^{d_f}$ where $L$ is a characteristic length and $d_f$
is the fractal dimension of the surface. Recently, the scaling behavior
of correlation functions in a critical system in two dimensions
near the fractal boundary of a random walk, 
for which $d_f = 4/3$, has been studied by methods 
of quantum gravity \cite{Dup98} and conformal invariance \cite{Car99}.
Another possibility is that
the substrate has a {\em self-affine\/} surface, for which the surface 
area is proportional to the projected area. In this case the height 
fluctuations are characterized by a roughness exponent $\zeta$ with 
$0 < \zeta < 1$, so that $(\delta h)^2 \sim L^{2 \zeta}$ where 
$\delta h$ is a typical height fluctuation over a distance $L$.
Self-affine scaling is predicted by many numerical and analytical 
models of surface growth \cite{BS95,Kar96}, and is also observed in a 
number of experiments \cite{KP95}.
A liquid-vapor interface, which exhibits rippled configurations 
due to the occurrence of capillary waves, is another
realization of a self-affine rough surface \cite{NPW88}.
An example where such an interface confines a critical system
is given by the interface between 
liquid $^4$He near the normalfluid-superfluid transition and its 
noncritical vapor, which occurs in a recently used experimental 
setup in which the Casimir force in a critical system is measured 
\cite{GC} (see also Ref.\,\cite{gold}).

In a previous Letter \cite{HK2001}, we showed that the shape of 
the surface has a distinct influence on the properties of an adjacent 
medium with long-range correlations. 
Here we demonstrate this in more detail for two-point correlation 
functions near a critical point of the medium, for both cases 
(i) and (ii) outlined above. Apart from Appendix \ref{appB},
we choose the Dirichlet boundary condition $\Phi = 0$
at the surface, which represents the so-called {\em ordinary} 
surface universality class in case of a flat surface, and is 
usually appropriate for magnets, binary alloys near a 
continuous order-disorder transition, and $^4$He near the 
normalfluid-superfluid transition \cite{Bin83,Die86}. 
In Ref.\,\cite{Pal}, the influence of surface roughness on the 
fluctuation properties of wetting films, and on the demagnetizing 
factor of a thin magnetic film, have been studied.  

In order to study the effects of the surface shape, 
we develop a perturbative expansion of two-point correlation 
functions in the deformations of the height profile. The method
is the path integral approach used previously to calculate
free energies \cite{LK91}, and in the context of the dynamic
\cite{GK97} and static \cite{EGHK2001} Casimir effect. 
Initially for a Gaussian field, the calculations are 
carried out to second order in the deformations. The first 
order results can also be derived by means of the stress tensor 
in conjunction with a novel type of short distance expansion 
(see Appendix \ref{appB}), and hold quite generally for any critical 
system bounded by a surface with either (a) Dirichlet boundary 
conditions $\Phi = 0$, or (b) boundary conditions that break the 
symmetry of the order parameter near the surface. In the latter 
case, the leading singular behavior can be obtained by setting 
$\Phi = \infty$ at the surface, corresponding to the 
{\em extraordinary} or {\em normal} surface universality class, 
describing {\em critical adsorption\/} of a binary liquid mixture on the 
surface of a substrate or the interface between the critical liquid 
and its noncritical vapor \cite{Bin83,Die86,Law2001}.
The second order results are particularly useful for cases in which 
the first order contributions vanish (see below).

The diffuse scattering of x-rays and neutrons at grazing incidence 
due to the modified correlations appears in addition to what would be
observed if the surface was separating two homogeneous media 
\cite{DH95}. The modified correlations may thus provide an additional 
and indirect means of characterizing the surface profile. This may 
be of value when other techniques are not possible, as in the case 
of the interior surface of a glass, or an internal crack, whereas 
scattering from a critical fluid or binary alloy coating the surface 
may be feasible.
Already at the first order, the two-point correlation functions
track the profile from the substrate, with a modulation that
decreases with the distance of the two points from the surface. 
This leads to explicit predictions for the structure factor, 
as a function of the lateral wave vector transfer, for a modulated 
surface. 

For self-affinely rough surfaces, second order calculations 
are necessary, as the first order results vanish on average.
In this context, the surface roughness is an example
of quenched randomness. 
For a massless Gaussian field, we find the expected result that
self-affine roughness leads to subleading corrections to the decay of 
two-point correlation functions, which at a scale $r$ are smaller
by a factor of $r^{-2(1-\zeta)}$ than the leading contribution 
coming from a flat surface.
Typical critical systems, however, are described 
by a non-Gaussian (interacting) field theory. In this case, the 
correlations are calculated perturbatively in a double expansion
in the deformations and in the strength of the 
interaction, and the results interpreted with the aid of the
renormalization group (RG) in $4 - \varepsilon$ dimensions.
We find that the subleading corrections now fall off with 
a slower power as compared to the Gaussian case
and, surprisingly, for a sufficiently large $\zeta$ even
{\em dominate\/}, giving rise to novel surface critical behavior.
However, for the XY model in two dimensions, below the 
Kosterlitz-Thouless temperature, we again find that the surface 
correlations fall off with the simple relative factor of 
$r^{-2(1-\zeta)}$ as compared to a flat surface (line). 

The results for correlation functions can also be related to 
thermodynamic quantities. To this end, we introduce 
distinct fields $h_b$ and $h_s$ in the bulk and close to the 
surface, respectively, and propose a scaling ansatz for the 
leading singular part of the surface free energy per projected 
area $f_s^{(\text{sing})}$. By taking suitable 
derivatives of $f_s^{(\text{sing})}$ with respect to $h_b$ and 
$h_s$, we then obtain scaling relations for a variety of critical 
exponents related to thermodynamic surface quantities.
 
The rest of the paper is organized as follows. 
In Sec.\,\ref{sectionfree} we introduce the geometry, and develop the
formalism for the perturbative calculation of correlation functions
for a free (Gaussian) field theory.
In Secs. \ref{sectionmod} and \ref{sectionrough} we then consider
a regularly patterned surface and a self-affinely rough surface 
in more detail. In Sec.\,\ref{sectionloop} we combine the
previous results with the RG, and obtain results for surface 
critical exponents. In Sec.\,\ref{sectionxy} we consider the 
XY model. Finally, in Sec.\,\ref{sectiondiscussion}, we draw 
our conclusions and outline some possible extensions of our
work; in particular, we relate our previous results for
correlation functions to thermodynamic 
surface quantities via scaling relations. Some technical details are 
left for Appendices \ref{appA} - \ref{appC}. 
In Appendix \ref{appB}, for instance, we introduce a new type 
of short-distance expansion for the stress tensor.

%%%%%%%%%%%%%%%%%%%%%%%%%%%%%%%%%%%%%%%%%%%%%%%%%%%%%%%%%%%%%%%%%

\section{Geometry and free field theory}
\label{sectionfree}

\begin{figure}[t]
\begin{picture}(16,8)
\put(0.6,1){
\setlength{\epsfysize}{7cm}
\epsfbox{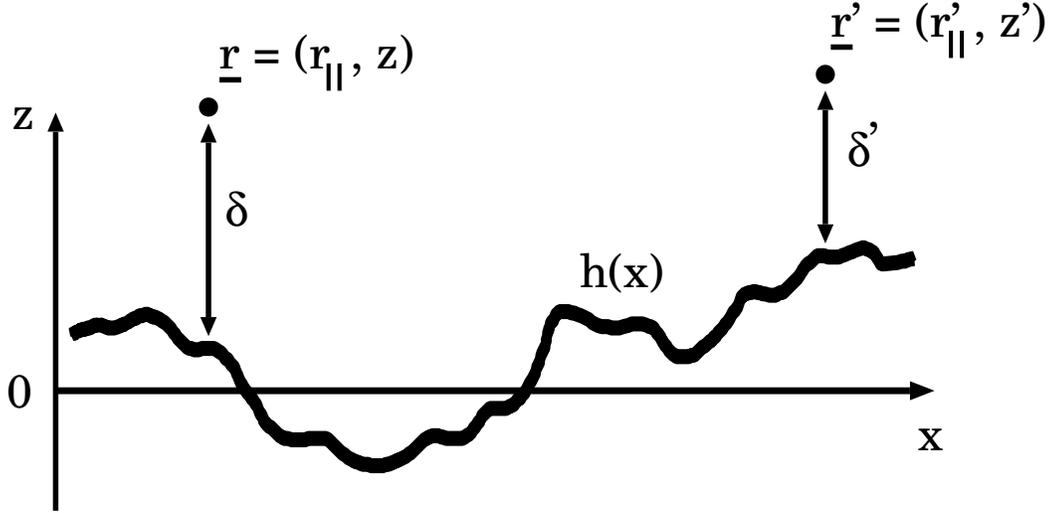}}
\end{picture}
\caption{Position vectors 
$\underline{r} = ({\bf r}_{\parallel}, z)$ and 
$\underline{r}' = ({\bf r}_{\parallel}', z')$ of the   
two-point correlation function in the critical system located 
above and bounded by a deformed surface. The surface profile is
described by the height function $h({\bf x})$, and the vertical
distances of $\underline{r}$ and $\underline{r}'$ from the surface 
are given by $\delta = z - h({\bf r}_{\parallel})$ and  
$\delta' = z' - h({\bf r}_{\parallel}')$, respectively.}
\label{fig1}
\end{figure}

We consider a manifold $\Omega$ with the shape of a deformed surface.
Each point on the manifold is represented by a vector
$X({\bf y}) = [X^{\mu}({\bf y}); \mu = 1, \ldots, d]$;
a $D$ dimensional manifold $\Omega$ embedded in $d$ dimensional 
space is parametrized by ${\bf y} = (y_1, \ldots, y_D)$. 
In the absence of overhangs and inlets, the surface 
profile can be described by a single-valued height function 
$h({\bf y})$, where ${\bf y}$ spans a $D = d - 1$ dimensional 
base plane (see Fig.\,\ref{fig1}). The parametrization of the
surface is thus $X({\bf y}) = [{\bf y}, h({\bf y})]$. 
Position vectors $\underline{r}$ are decomposed according 
to $\underline{r} = ({\bf r}_{\parallel}, z)$, where 
${\bf r}_{\parallel}$ comprises the $D = d - 1$ components 
parallel to the surface, and $z$ is the distance from the 
base plane. The vertical distance of $\underline{r}$ from 
the surface is given by $\delta = z - h({\bf r}_{\parallel})$
(see Fig.\,\ref{fig1}). 
We denote $d$ dimensional vectors with underlined letters, 
and $D$ dimensional vectors with boldface letters.

Fluctuations in the critical system located above the surface 
will be described by an $n$-component order parameter field 
$\Phi(\underline{r}) = 
[\Phi_1(\underline{r}), \ldots , \Phi_n(\underline{r})]$.
We consider the statistical Boltzmann weight 
$e^{- \beta {\cal H}}$ with standard Hamiltonian 
\cite{Bin83,Die86}
\begin{equation} \label{hamiltonian}
\beta {\cal H}\{ \Phi \} \, = \,   
\int_V d^d r \, \left\{ \frac{1}{2} (\nabla \Phi)^2 \, + \,
\frac{\tau_0}{2} \, \Phi^2 
\, + \, \frac{u_0}{4!} \, (\Phi^2)^2 \right\} \, \, ,
\end{equation}

%%%%%%%%%%%%%%%%%%%%%%%% Fig.: parallel.ps %%%%%%%%%%%%%%%%%%%%%%%%%
%
\unitlength1cm
%
%%%%%%%%%%%%%%%%%%%%%%%%%%%%%%%%%%%%%%%%%%%%%%%%%%%%%%%%%%%%%%%%%%%%

\noindent
where $\tau_0 \sim T - T_c$ and $u_0$ is the strength of the 
$\Phi^4$ interaction. In this Section, we study the Gaussian 
theory, for which $u_0 = 0$. 
The volume $V$ consists of the space available to the critical 
system. The above expression must be supplemented by a boundary 
condition on the surface. 
We choose the Dirichlet boundary condition $\Phi = 0$,
representing the ordinary surface universality class. 
In this case, for $n = 1$ the order parameter $\Phi$ can represent 
the magnetization in a uniaxial ferromagnet or the deviation of the 
composition in a binary alloy from the critical composition, for 
$n = 2$ the magnetization in a XY-magnet or the superfluid order 
parameter of $^4$He near the normalfluid-superfluid transition, 
and for $n = 3$ the magnetization in a Heisenberg ferromagnet
\cite{Bin83,Die86}.

The Gaussian two-point correlation function 
(or {\em propagator\/})
\begin{equation} \label{gauss}
\langle \Phi_i(\underline{r}) \Phi_j(\underline{r}') \rangle \, = \, 
\delta_{ij} \, G(\underline{r} ; \underline{r}') \, \, , 
\quad u_0 = 0 \, \, ,
\end{equation}
where the brackets $\langle \, \, \, \, \rangle$ 
denote the thermal average according to  
Eq.\,(\ref{hamiltonian}) with $u_0 = 0$,
can be calculated using functional integral methods 
\cite{LK91,GK97}. The details of this calculation are
left to Appendix \ref{appA}. The result is 
\begin{eqnarray}
G(\underline{r}; \underline{r}') \, & = & \, 
G_b(\underline{r}; \underline{r}') \, -
\int d^D x \int d^D y  \label{gauss2} \\
& & \times \, G_b[\underline{r} ; {\bf x}, h({\bf x})] \,
M({\bf x}, {\bf y}) \, G_b[\underline{r}' ; {\bf y}, h({\bf y})] 
\, \, , \nonumber
\end{eqnarray}
where 
\begin{equation} \label{bulk}
G_b(\underline{r}; \underline{r}') \, = \, 
\int \frac{d^D p}{(2 \pi)^D} \, 
e^{i {\bf p} \cdot ({\bf r}_{\parallel} - {\bf r}_{\parallel}' ) }
\, \frac{1}{2 p} \, e^{- p |z - z'|} 
,
\end{equation}
with $p = |{\bf p}|$, is the Gaussian propagator in unbounded bulk, 
and the kernel $M({\bf x}, {\bf y})$ is the inverse of the kernel 
$G_b[{\bf x}, h({\bf x}); {\bf y}, h({\bf y})]$, i.e.
\begin{equation} \label{inverse}
\int d^D y \, M({\bf x}, {\bf y}) \, 
G_b[{\bf y}, h({\bf y}); {\bf y}', h({\bf y}')] \, = \,
\delta^D({\bf x} - {\bf y}') \, .
\end{equation}
While the above results [with an appropriately modified bulk 
propagator in Eq.(\ref{bulk})]
are generally valid, we focus on the 
behavior of the correlation functions at the bulk critical point, 
i.e., for $T = T_c$, where correlations are strongest 
\cite{HSED98}.

Equation (\ref{gauss2}) is difficult to evaluate in general. 
To proceed, we now consider the height profile $h({\bf x})$ 
as a small perturbation, and expand 
$G(\underline{r} ; \underline{r}')$ in a series 
$G_0 + G_1 + G_2 + \ldots$ in powers of $h$ up 
to second order, under the constraint that 
$z$ and $z'$ are kept fixed. The lowest order result,
\begin{eqnarray} \label{h0z}
G_0(\underline{r} ; \underline{r}') \, & = & \,  
G_b({\bf r}_{\parallel} , z ; {\bf r}_{\parallel}', z') -
G_b({\bf r}_{\parallel} , z ; {\bf r}_{\parallel}', - z') \\[1mm]
& = & \, \int \frac{d^D p}{(2 \pi)^D} \, 
e^{i {\bf p} \cdot ({\bf r}_{\parallel} - {\bf r}_{\parallel}' ) } \, 
g_0(p ; z, z')
\end{eqnarray}
with [see Eq.\,(\ref{bulk})]
\begin{equation} \label{half2}
g_0(p ; z, z') \, = \,
\frac{1}{2 p} \, 
\Big[ e^{- p |z - z'|} 
- e^{- p (z + z')} \Big]
\end{equation}
corresponds to a flat surface, and can be obtained by the method
of images \cite{Bin83,Die86}. The bulk correlation function
$G_b(\underline{r} ; \underline{r}')$ decays as $r^{-(d-2+\eta)}$
for large separations $r = |\underline{r} - \underline{r}'|$, 
where the bulk critical exponent $\eta$ is given by
$\eta = 0$ in the Gaussian theory.
In contrast, if both points remain
close to the surface, $G_0(\underline{r} ; \underline{r}')$ decays as 
$r^{-(d-2+\eta_{\parallel})}$, where $\eta_{\parallel}$
is a surface critical exponent given by 
$\eta_{\parallel} = 2$ in the Gaussian theory 
\cite{Bin83,Die86}.

The first order result is given by \cite{MG99}
\begin{equation} \label{h1z}
G_1(\underline{r} ; \underline{r}') \, = \, - \, 4
\int d^D x \, J({\bf r}_{\parallel}, {\bf x} ; z)
\, h({\bf x})
J({\bf r}_{\parallel}', {\bf x} ; z') , \nonumber
\end{equation}
with
\begin{equation} \label{J}
J({\bf x}, {\bf y} ; z) \, = \, \frac{1}{2}
\int \frac{d^D p}{(2 \pi)^D} \,
e^{i {\bf p} \cdot ({\bf x} - {\bf y})} \,
e^{- p z} \, \, .
\end{equation}
Note that $J({\bf x}, {\bf y}; z \to 0^+) 
= \frac{1}{2} \delta^D({\bf x} - {\bf y})$, where
$\delta^D({\bf x})$ is the delta function in $D$ dimensions.
Already the result at this order tracks the profile 
$h({\bf x})$ of the surface. For example, for 
$\rho = |{\bf r}_{\parallel} - {\bf r}_{\parallel}'| \to \infty$
with $z$ and $z'$ fixed, the above results for $G_0$ and $G_1$ 
imply the behavior (see Appendix \ref{appB})
\begin{equation} \label{leading}
G(\underline{r} ; \underline{r}') \sim  
[ 1 - A(\underline{r}) - A(\underline{r}') ] \, 
\rho^{- (d - 2 + \eta_{\parallel})} \, \, ,
\end{equation}
up to terms of order $(h/z)^2$ and $(h/z')^2$. 
Thus, the leading power law is the same as for a flat surface,
but the amplitude is modulated by the surface deformations in the 
vicinity of ${\bf r}_{\parallel}$ and ${\bf r}_{\parallel}'$ by
the dimensionless and universal amplitude
\begin{equation} \label{A1}
A(\underline{r}) \, = \,
\frac{\eta_{\parallel} - \eta}{2} \,
\int d^D x \, \frac{h({\bf x})}{z} \,
\Delta( {\bf x} - {\bf r}_{\parallel}, z) \, \, ,
\end{equation}
where for the present Gaussian case, $\eta_{\parallel} - \eta = 2$
and $\Delta({\bf x} - {\bf r}_{\parallel}, z) = 
2 J({\bf x}, {\bf r}_{\parallel}; z)$.
Equations (\ref{leading}) and (\ref{A1}) are valid quite generally, 
and in particular also for the boundary condition representing 
critical adsorption of a binary liquid mixture 
(see Appendix \ref{appB}).
The explicit form of $\Delta({\bf x}, z)$, however, 
depends on the surface universality class considered. 

The second order result reads
\begin{equation} \label{h2z}
G_2(\underline{r} ; \underline{r}') \, = \, 
\int d^D x \int d^D y \, h({\bf x}) h({\bf y}) \,
C(\underline{r} , \underline{r}'; {\bf x}, {\bf y})
,
\end{equation}
with
\begin{equation} \label{zG2K} 
C(\underline{r} , \underline{r}'; {\bf x}, {\bf y}) \, = \, 
- \, 8 \, J({\bf r}_{\parallel}, {\bf x} ; z) \,
J({\bf r}_{\parallel}', {\bf y} ; z') \,
K({\bf x}, {\bf y} ; z \to 0^+)
,
\end{equation}
and
\begin{equation} \label{K} 
K({\bf x}, {\bf y}; z) \, = \, \frac{1}{2}
\int \frac{d^D p}{(2 \pi)^D} \,
e^{i {\bf p} \cdot ({\bf x} - {\bf y})} \, 
p \, e^{- p z} \, \, .
\end{equation}

In a scattering experiment with grazing incidence, one probes the
lateral structure factor $S({\bf p}, z; {\bf p}',z')$
\cite{DW84,Dosch92,DH95}, which is defined by the Fourier transform
\begin{equation} \label{sf}
G(\underline{r} ; \underline{r}') \, = \,
\int \frac{d^D p}{(2 \pi)^D} \,
e^{i {\bf p} \cdot {\bf r}_{\parallel}} 
\int \frac{d^D p'}{(2 \pi)^D} \,
e^{i {\bf p}' \cdot {\bf r}_{\parallel}'} \, 
S({\bf p}, z; {\bf p}',z') \, \, .
\end{equation}
Using the Fourier transform of the height profile
\begin{equation} \label{hk}
h({\bf y}) = \int \frac{d^D k}{(2 \pi)^D} \,
e^{i {\bf k} \cdot {\bf y}} \, \hat{h}({\bf k})
,
\end{equation}
with $\hat{h}(-{\bf k}) = \hat{h}({\bf k})^{\ast}$, we obtain
an equivalent expansion $S = S_0 + S_1 + S_2 + \ldots$, with
\begin{eqnarray}
S_0 \, & = & \, \frac{1}{2 p} \, \left[
e^{- p |z - z'|} - e^{- p |z + z'|} \right] \,
(2 \pi)^D \, \delta({\bf p} + {\bf p}') \, \, , \label{s0}\\[2mm]
S_1 \, & = & \, - \, 
e^{- p z} \, e^{- p' z'} \, \hat{h}({\bf p} + {\bf p}') \, \, , 
\label{s1}\\[2mm]
S_2 & = & - \, 
e^{- p z} \, e^{- p' z'}
\int \frac{d^D k}{(2 \pi)^D} \, |{\bf p} - {\bf k}| \,
\hat{h}({\bf k}) \, \hat{h}({\bf p} + {\bf p}' - {\bf k}) 
\, \, . \label{s2}
\end{eqnarray}

For a rough surface, the deviations in height
from a planar surface have no upper bound. 
In this case, it is convenient to carry out the 
expansion in $h({\bf x})$ for fixed vertical distances 
$\delta  = z - h({\bf r}_{\parallel})$ and
$\delta' = z - h({\bf r}_{\parallel}')$, instead of for
fixed $z$ and $z'$ (see Fig.\,\ref{fig1}).
This representation facilitates the perturbative analysis of the 
field theory described by Eq.\,(\ref{hamiltonian}) 
(see Sec.\,\ref{sectionloop}).
Moreover, in view of probing correlation functions 
lateral to the substrate surface by grazing incidence scattering of 
x-rays and neutrons \cite{DW84,Dosch92,DH95}, 
this representation is natural,
since in these experiments $\delta$ and $\delta'$ show up as
length scales which are set by the finite penetration depth of 
the x-rays.

Writing $G = G_0 + G_{\text{I}} + G_{\text{II}} + \ldots$ where the
subscripts 0, I, II indicate the corresponding order in $h({\bf x})$
under the constraint that $\delta$ and $\delta'$ are kept fixed,
we find
\begin{equation} \label{h0delta}
G_0(\underline{r} ; \underline{r}') \, = \, 
G_b({\bf r}_{\parallel} , \delta ; {\bf r}_{\parallel}', \delta') \, - \,
G_b({\bf r}_{\parallel} , \delta ; {\bf r}_{\parallel}', - \delta') \, \, , 
\end{equation}
\begin{eqnarray} \label{h1delta}
G_{\text{I}}(\underline{r} ; \underline{r}') \, & = & \, 
- \, [h({\bf r}_{\parallel}) - h({\bf r}_{\parallel}')] \, \, 
\frac{\partial}{\partial \delta'} \, 
G_b({\bf r}_{\parallel}, \delta ; {\bf r}_{\parallel}' , \delta') \\[1mm]
& & + \, 2
\int d^D x \, \, J({\bf r}_{\parallel}, {\bf x} ; \delta) \,
[h({\bf r}_{\parallel}) + h({\bf r}_{\parallel}') - 2 h({\bf x})] \, 
J({\bf r}_{\parallel}', {\bf x} ; \delta')
\, \, , \nonumber
\end{eqnarray}
\begin{eqnarray} \label{h2delta}
G_{\text{II}}(\underline{r} ; \underline{r}') \, & = & \, 
\frac{1}{2} \, \Big[ 
K({\bf r}_{\parallel}, {\bf r}_{\parallel}' ; |\delta - \delta'| ) +
K({\bf r}_{\parallel}, {\bf r}_{\parallel}' ;  \delta + \delta'  ) 
\Big] \, \, 
[h({\bf r}_{\parallel}) - h({\bf r}_{\parallel}')]^2  \\[1mm]
& & + \int d^D x \int d^D y \, \, 
J({\bf r}_{\parallel}, {\bf x} ; \delta) \,
M_0({\bf x}, {\bf y}) \, [h({\bf x}) - h({\bf y})]^2 \,
J({\bf r}_{\parallel}' , {\bf y} ; \delta') \nonumber  \\[1mm]
& & - \, 2 \left[ \, \int d^D x \, \,
K({\bf r}_{\parallel}, {\bf x} ; \delta) \,
[h({\bf r}_{\parallel}) - h({\bf x})]^2 \, 
J({\bf r}_{\parallel}' , {\bf x} ; \delta')
\, \, + \, ({\bf r} \leftrightarrow {\bf r}') \, \right] \, \, . \nonumber
\end{eqnarray}
The first line in Eq.\,(\ref{h1delta}) is valid for $\delta' < \delta$,
and $M_0({\bf x}, {\bf y})$ in Eq.\,(\ref{h2delta}) is defined as in
Eq.\,(\ref{inverse}) but with $h({\bf y}) = 0$. 
The kernels $J$ and $K$ are given by
Eqs.\,(\ref{J}) and (\ref{K}), respectively.
The contribution $G_0$ in Eq.\,(\ref{h0delta}) corresponds to the 
Gaussian propagator for a half-space bounded by a flat surface
with Dirichlet boundary conditions, i.e.,
\begin{equation} \label{halfdelta}
G_0(\underline{r} ; \underline{r}') \, = \, 
\int \frac{d^D p}{(2 \pi)^D} \, 
e^{i {\bf p} \cdot ({\bf r}_{\parallel} - {\bf r}_{\parallel}' ) } \, 
g_0(p ; \delta, \delta')
\end{equation}
with $g_0$ from Eq.\,(\ref{half2}).

%%%%%%%%%%%%%%%%%%%%%%%%%%%%%%%%%%%%%%%%%%%%%%%%%%%%%%%%%%%%%%

\section{Modulated surfaces}
\label{sectionmod}

We now apply the results of the previous Section to
patterned surfaces. The simplest example is provided 
by an uniaxial sinusoidal modulation with wavelength
$\lambda$ along, say, the $x$ direction, i.e., 
\begin{equation} \label{mod}
h(x,{\bf Y}) \, = \, a \cos(2 \pi x / \lambda) \, \, .
\end{equation}
The other $D-1$ directions along the surface, denoted by 
${\bf Y}$, remain translationally invariant. 
The Fourier transform of this height profile is
\begin{equation} \label{modf}
\hat{h}({\bf k}) \, = \, \frac{a}{2} \,
(2 \pi)^D \, \delta^{D-1}({\bf K}) \,
\left[\delta(k_x - \textstyle{\frac{2 \pi}{\lambda}}) +
      \delta(k_x + \textstyle{\frac{2 \pi}{\lambda}}) \right] \, \, ,
\end{equation}
where ${\bf k}$ is decomposed according to 
${\bf k} = (k_x,{\bf K})$.

The nontrivial orders of the expansion of 
$G(\underline{r} ; \underline{r}')$ in $h$ 
for fixed $z$ and $z'$ are given by
\begin{eqnarray} \label{modg1}
G_1({\bf r}_{\parallel}, z; {\bf r}_{\parallel}' , z') \, & = & \,
- \, \frac{a}{2} \, 
e^{\frac{2 \pi i}{\lambda} x'}
\int \frac{d^D p}{(2 \pi)^D} \,
e^{i {\bf p} \cdot ({\bf r}_{\parallel} - {\bf r}_{\parallel}')} \\[1mm]
& & \quad \times \, 
e^{- p z} \,
e^{- |{\bf p} - (\frac{2 \pi}{\lambda},0)| z' } \quad
+ \, ({\bf r}_{\parallel} \leftrightarrow {\bf r}_{\parallel}') 
\, \, , \nonumber
\end{eqnarray}
\begin{eqnarray} \label{modg2}
G_2({\bf r}_{\parallel}, z; {\bf r}_{\parallel}' , z') \, & = & \,
- \, \frac{a^2}{4} \, 
\int \frac{d^D p}{(2 \pi)^D} \,
e^{i {\bf p} \cdot ({\bf r}_{\parallel} - {\bf r}_{\parallel}')} \,
|{\bf p} - (\textstyle{\frac{2 \pi}{\lambda}},0)| \,
e^{- p z} \, e^{- p z' } \\[1mm] 
& & \, - \, \frac{a^2}{4} \, 
e^{\frac{4 \pi i}{\lambda} x'} \,
\int \frac{d^D p}{(2 \pi)^D} \,
e^{i {\bf p} \cdot ({\bf r}_{\parallel} - {\bf r}_{\parallel}')} 
\nonumber \\[1mm]
& & \quad \times \,
|{\bf p} - (\textstyle{\frac{2 \pi}{\lambda}},0)| \,
e^{- p z} \,
e^{- |{\bf p} - (\frac{4 \pi}{\lambda},0)| z' } \quad
+ \, ({\bf r}_{\parallel} \leftrightarrow {\bf r}_{\parallel}') 
\, \, . \nonumber
\end{eqnarray}
For $\rho = |{\bf r}_{\parallel} - {\bf r}_{\parallel}'| \to \infty$,
the leading power law in $\rho$ is the same as for a flat surface,
but the amplitude is modulated by the shape of the surface in the 
vicinity of ${\bf r}_{\parallel}$ and ${\bf r}_{\parallel}'$. 
In particular, the first order result in Eq.\,(\ref{modg1}) 
is consistent with Eqs.\,(\ref{leading}) and (\ref{A1}).
For $z, z' \ll a, \lambda$ the correlations follow more or
less the surface modulation. Interestingly, for 
$z, z' \gg \lambda$, the correlations that are sensitive 
to the modulation, i.e., depend on $\lambda$, decay 
{\em exponentially\/} in $z/\lambda$.
For instance, for $z = z'$ and $z / \lambda \to \infty$, one
has $G_1 \sim e^{- \frac{2 \pi}{\lambda} z}$ and
$G_2 \sim e^{- \frac{4 \pi}{\lambda} z}$.
This exponential decay is due to the fact that the 
surface profile (\ref{mod}) has a perfect periodic shape. 
In contrast, a {\em local\/} perturbation on the surface would 
result in a perturbation of the correlations which decays only
as a power law with the distance from the surface.  

\newpage

\noindent
The corresponding orders of the lateral structure factor 
are given by
\begin{eqnarray} \label{mods1}
S_1({\bf p}, z; {\bf p}',z') \, & = & \,
- \, \frac{a}{2} \, e^{- p z} \, e^{- p' z'} \,
(2 \pi)^D \, \delta^{D-1}({\bf P} + {\bf P}') \\[1mm]
& & \, \times
\left[\delta(p_x + p_x' - \textstyle{\frac{2 \pi}{\lambda}}) +
      \delta(p_x + p_x' + \textstyle{\frac{2 \pi}{\lambda}}) \right]
\, \, , \nonumber
\end{eqnarray}
\begin{eqnarray} \label{mods2}
S_2({\bf p}, z; {\bf p}',z') & = & 
- \, \frac{a^2}{4} \, \, e^{- p z} \, e^{- p' z'} \,
(2 \pi)^D \, \delta^{D-1}({\bf P} + {\bf P}') \\[1mm]
& & \times \, \Big\{
|{\bf p} - (\textstyle{\frac{2 \pi}{\lambda}},0)| 
\left[\delta(p_x + p_x' - \textstyle{\frac{4 \pi}{\lambda}}) +
      \delta(p_x + p_x') \right] \nonumber \\[1mm]
& & \, \, \, \, + \, 
|{\bf p} + (\textstyle{\frac{2 \pi}{\lambda}},0)| 
\left[\delta(p_x + p_x' + \textstyle{\frac{4 \pi}{\lambda}}) +
      \delta(p_x + p_x') \right] \Big\} \, \, . \nonumber
\end{eqnarray}
These results indirectly characterize the surface in scattering 
experiments. For instance, the form of $S_1$ implies that the 
incident wave vector component $p_x$ is scattered to 
$p_x' = p_x \pm \frac{2 \pi}{\lambda}$ while the other 
components of ${\bf p}$ remain unchanged. 
The form of $S_2$ implies that 
$p_x$ is scattered by $\frac{4 \pi}{\lambda}$, $0$, 
$- \frac{4 \pi}{\lambda}$. 
In a scattering experiment with grazing incidence, the length 
scale perpendicular to the surface is set by the depth $b$
that the evanescent wave penetrates the sample, giving rise to
diffuse scattering and thereby probing the critical correlations 
close to the surface \cite{DW84}. Since this diffuse scattering 
appears in addition to the contribution already present away
from criticality \cite{DH95}, it can in principle be separated 
out by tuning the temperature deviation $T - T_c$. 
We assume that $b$ is much larger than the 
height of the deformations.
In this case, the above expansion in the deformations results
in an expansion in powers of $h / b \ll 1$ for the 
elastic scattering cross section, which 
allows one to distinguish the corresponding contributions
via their intensities.

%%%%%%%%%%%%%%%%%%%%%%%%%%%%%%%%%%%%%%%%%%%%%%%%%%%%%%%%%%%%%%

\section{Rough surfaces}
\label{sectionrough}

The second order results are particularly useful when dealing
with rough surfaces, where the quench averaged first order corrections
vanish. Within the description using a height function $h({\bf x})$,
self-affine roughness is characterized by the behavior
\begin{equation} \label{zeta}
\overline{ [h({\bf x}) - h({\bf y})]^2 } \, \sim \, \, 
|{\bf x} - {\bf y}|^{2 \zeta} \, \, , \quad
|{\bf x} - {\bf y}| \to \infty \, \, ,
\end{equation}
where the overbar denotes averaging over self-affine realizations
of the surface profile, and $\zeta$ with $0 < \zeta < 1$ is the 
roughness exponent. Without restriction of the generality we choose 
the coordinate system so that $\overline{ h(\bf x) } = 0$.
In the limit of short distances $|{\bf x} - {\bf y}|$
it is reasonable to assume that the surface is smooth.
This can be modelled by the Fourier transform
\begin{eqnarray}
\overline{ [h({\bf x}) - h({\bf y})]^2 } \, & = & \, 
\omega^{2 - 2 \zeta} \, |{\bf x} - {\bf y}|^2 \label{crossover}\\
& & \times \int \frac{d^D p}{(2 \pi)^D} \, 
e^{i {\bf p} \cdot ({\bf x} - {\bf y} ) } \, 
p^{-D + 2 - 2 \zeta} \, e^{-p \lambda} \, \, . \nonumber
\end{eqnarray}
While at large separations the above correlations grow as 
$|{\bf x} - {\bf y}|^{2 \zeta}$, we have also introduced a 
cutoff length $\lambda$ to regulate the behavior of the surface 
at short distances, and an overall amplitude length $\omega$. 
The length $\lambda$ characterizes the crossover from the analytic 
behavior for $|{\bf x} - {\bf y}| \ll \lambda$ to the behavior in 
Eq.\,({\ref{zeta}) for $|{\bf x} - {\bf y}| \gg \lambda$. Apart from
its physical significance, the appearance of the finite crossover length 
$\lambda$ in Eq.\,(\ref{crossover}) is also essential within the present 
theoretical approach (see Sec.\,\ref{sectionloop}). 

A characteristic feature of self-affine roughness is statistical
translational
invariance, since the right hand side of Eq.\,(\ref{crossover}) depends
on the distance $|{\bf x} - {\bf y}|$ only. 
This implies that the averaged lateral structure factor 
$\overline{S}$ is proportional to
$\delta^D({\bf p} + {\bf p}')$, and depends on 
$z$, $z'$, and $p$ = $|{\bf p}|$ only.
In order to maintain translational invariance, 
it is convenient to express the 
results for the correlation functions in terms of the local distance 
$\delta = z - h({\bf r}_{\parallel})$ from the surface rather than $z$
(see Fig.\,\ref{fig1}).
The two-point correlation function must now vanish as
$\delta$ or $\delta'$ go to zero.
On averaging $G(\underline{r} ; \underline{r}')$ over different 
surface profiles, the contribution $G_{\text{I}}$ in Eq.\,(\ref{h1delta}) 
vanishes due to $\overline{ h({\bf x}) } = 0$, and the contribution 
$G_{\text{II}}$ in Eq.\,(\ref{h2delta}) becomes translationally 
invariant with respect to the lateral components 
${\bf r}_{\parallel}$ and ${\bf r}_{\parallel}'$. 
We thus introduce the lateral Fourier transform
\begin{equation} \label{g}
\overline{G_{\text{II}}(\underline{r} ; \underline{r}')} \, = \, 
\int \frac{d^D p}{(2 \pi)^D} \, 
e^{i {\bf p} \cdot ({\bf r}_{\parallel} - {\bf r}_{\parallel}' ) } \, 
g_2(p; \delta, \delta') \, \, ,
\end{equation}
where $g_2(p; \delta, \delta')$ can be read off from the 
right hand side of Eq.\,(\ref{h2delta}), i.e.,
\begin{eqnarray} 
g_2(p; \delta, \delta') \, & = & \, 
\frac{1}{2} \, 
\Big[ \, {\cal K}(p, |\delta - \delta'|) +
{\cal K}(p, \delta + \delta') \, \Big] \label{G2}\\[1mm]
& & + \, {\cal K}(p, 0) \, e^{-p (\delta + \delta')} \, - \,
{\cal K}(p, \delta) \, e^{-p \delta'} \, - \,
{\cal K}(p, \delta') \, e^{-p \delta} \, \, . \nonumber
\end{eqnarray}
${\cal K}(p,\delta)$ is the lateral Fourier transform of
$K({\bf x}, {\bf y} ; \delta) \, \,
\overline{ [h({\bf x}) - h({\bf y})]^2 }$ and we have used 
the fact that the lateral Fourier transform of 
$M_0({\bf x}, {\bf y}) \, \overline{ [h({\bf x}) - h({\bf y})]^2 }$
appearing in the second line of Eq.\,(\ref{h2delta}) after averaging
is given by $4 \, {\cal K}(p,\delta = 0)$. 
Using Eq.\,(\ref{crossover}),
${\cal K}(p,\delta)$ can be expressed in
terms of the convolution integral
\begin{equation} \label{Gp}
{\cal K}(p,\delta) \, = \, \omega^{2 - 2 \zeta} 
\int \frac{d^D k}{(2 \pi)^D} \, U(|{\bf p} - {\bf k}|, \delta) \,
k^{-D + 2 - 2 \zeta} \, e^{-k \lambda} \, \, ,
\end{equation}
where $U(p, \delta)$ is the lateral Fourier transform of 
$K({\bf x}, {\bf y}; \delta) \, |{\bf x} - {\bf y}|^2$
given by
\begin{equation} \label{Up}
U(p, \delta) \, = \, \left[
\delta \, - \, \frac{1}{2} \, p \, \delta^2 \, - \,
\frac{D-1}{2} \left( \frac{1}{p} \, - \, \delta \, \right)
\right] e^{- p \delta} \, \, .
\end{equation}
In terms of the coordinates 
$\underline{r} = ({\bf r}_{\parallel},\delta)$, the above results 
imply that the leading power law behavior of 
$\overline{G(\underline{r} ; \underline{r}')}$ for
$\rho = |{\bf r}_{\parallel} - {\bf r}_{\parallel}'| \to \infty$
is the same as for a flat surface. The corresponding amplitude 
depends on the roughness, and is modified by a factor of
$[1 - \kappa \, (\omega / \lambda)^{2(1-\zeta)}]$ as compared to
a flat surface, where $\kappa > 0$ is a number of order unity.
The subleading correction of order $\overline{h^2}$ decays with 
the separation $\rho$ with an additional factor of 
$\rho^{-2(1-\zeta)}$ compared to the leading term 
[see Eqs.\,(\ref{para}) and (\ref{etaparalleltilde}) in
Sec.\,\ref{sectionloop} for $\varepsilon = 0$].

Note that $g_2(p; \delta, \delta')$ vanishes for 
$\delta = 0$ or $\delta' = 0$ as it should, according 
to the Dirichlet boundary condition at the surface.
This would {\em not\/} be the case for $z = 0$ or $z' = 0$ 
if we carried out the expansion in $h({\bf x})$ 
with fixed $z$ and $z'$. However, the realization 
of the Dirichlet boundary condition for the Gaussian propagator
is essential for the perturbation
theory of the field theory described by Eq.\,(\ref{hamiltonian}).
Moreover, $g_2(p; \delta, \delta')$ is an analytic function for small 
$\delta$ or $\delta'$ due to the finite crossover length $\lambda$ 
in Eq.\,(\ref{Gp}), which would otherwise be ill defined for $\lambda = 0$ 
if $\delta = 0$ and $\zeta < 1/2$.

%%%%%%%%%%%%%%%%%%%%%%%%%%%%%%%%%%%%%%%%%%%%%%%%%%%%%%%%%%%%%%%%%%%%

\section{Interacting theory}
\label{sectionloop}

In this Section we consider the asymptotic scaling behavior of 
the two-point correlation function near a self-affine rough 
surface for the $n$-vector model at the bulk critical point. 
By combining our previous results with 
the field-theoretic renormalization group (RG), we estimate 
the values of the corresponding critical exponents,
using a double expansion in the surface deformations and 
in the deviation $\varepsilon = 4 - d$ of the space dimension $d$ 
from the upper critical dimension. 

For the interacting field theory, governed by Eq.\,(\ref{hamiltonian})
with $u_0 \neq 0$, standard perturbation theory can be applied to get
the correlation function near a surface of arbitrary but fixed shape,
\begin{equation} \label{loop1}
\langle \Phi_i(\underline{r}) \Phi_j(\underline{r}') \rangle \, = \, 
\delta_{ij} \, {\cal G}(\underline{r} , \underline{r}' ; u_0)
,
\end{equation}
with
\begin{equation} \label{loop2}
{\cal G}(\underline{r} , \underline{r}' ; u_0) \, = \, 
G(\underline{r} ; \underline{r}') \, - \,
\frac{n + 2}{3} \, \frac{u_0}{2} \,
\int_V d^d R \, \, 
G(\underline{r} ; \underline{R}) \,
G(\underline{R} ; \underline{R}) \, G(\underline{R} ; \underline{r}')
\, \, + \, {\cal O}(u_0^2) \, \, ,
\end{equation}
where the Gaussian propagator $G(\underline{r} ; \underline{r}')$ 
is given by Eq.\,(\ref{gauss2}). 
We are interested in the behavior of 
$\langle \Phi_i(\underline{r}) \Phi_j(\underline{r}') \rangle$ 
in the limit for which
the distance between $\underline{r}$ and $\underline{r}'$ is much larger than
one or both of the vertical distances 
$\delta$ and $\delta'$ (see Fig.\,\ref{fig1}).
If $\delta'$ is small, say, 
it is helpful to consider the so-called {\em surface operator\/} 
\cite{Bin83,Die86}
\begin{equation} \label{sop}
\Phi^{\perp}({\bf r}_{\parallel}') \, \equiv \, 
\partial_n \Phi(\underline{r}') \, \, ,
\end{equation}
where $\partial_n = [g({\bf r}_{\parallel}')]^{-1/2} 
[\partial_{\delta'} - \nabla h({\bf r}_{\parallel}') \cdot \nabla]$
denotes the normal derivative at ${\bf r}_{\parallel}'$ on the surface,
with the determinant 
$g({\bf y}) = 1 + [\nabla h({\bf y})]^2$
of its induced metric [see Eq.\,(\ref{metric})].
In this way one avoids to deal with the irrelevant length $\delta'$ 
from the outset. For correlations vertically away from the
surface, i.e., ${\bf r}_{\parallel} = {\bf r}_{\parallel}'$,
we are thus led to consider
\begin{equation} \label{Fperp}
\langle \Phi_i(\underline{r}) 
\Phi_j^{\perp}({\bf r}_{\parallel}) 
\rangle \, = \, \delta_{ij} \, 
{\cal G}_{\perp}({\bf r}_{\parallel} , \delta ; u_0) \, \, .
\end{equation}
The loop expansion of 
${\cal G}_{\perp}({\bf r}_{\parallel} , \delta ; u_0)$ 
is obtained by taking the normal derivative at ${\bf r}_{\parallel}$
of the right hand side of Eq.\,(\ref{loop2}). 

Up to now in this Section we have considered a surface with arbitrary 
but fixed shape. In particular,
for a {\em flat\/} surface, the one-loop addition in $u_0$ can 
be regularized and renormalized by minimal subtraction of 
poles in $\varepsilon = 4 - d$, leading to logarithmic
contributions in the separation $r = |\underline{r} - \underline{r}'|$.
This perturbative result can then be improved by RG,
resulting in power laws in $r$
with corresponding surface critical exponents \cite{Bin83,Die86}.
For a self-affinely rough surface, 
the function ${\cal G}_{\perp}$ depends,
of course, on the shape of this surface, i.e., on the 
height function $h({\bf x})$. However, by averaging over
different surface profiles, we expect that the average
$\overline{ {\cal G}_{\perp} }$ depends 
only on gross features characterizing the surface configurations,
and in particular becomes independent of ${\bf r}_{\parallel}$ 
due to translational invariance.
In the following we restrict ourselves to surfaces which are
rough on large distances, and to contributions 
to ${\cal G}_{\perp}$ up to second
order in $h({\bf x})$. According to Eq.\,(\ref{crossover})
we conclude that in this case the amplitude $\omega$ and the 
crossover length $\lambda$ are the only remaining
relevant length scales characterizing the different surface
configurations.

In the next step, the resulting average 
$\overline{ {\cal G}_{\perp}}(\delta; \omega, \lambda; u_0)$ 
has to be renormalized. For our perturbative calculations
we use dimensional regularization and renormalization by minimal
subtraction of poles in $\varepsilon = 4 - d$ \cite{ZJ89}.
The reparametrizations
\begin{equation} \label{uren}
u_0 \, = 16 \pi^2 \mu^{\varepsilon} Z_u u
\end{equation}
and 
\begin{equation} \label{phiren}
\Phi \, = \, Z_{\Phi}^{1/2} \, \Phi_{ren}
\end{equation}
of the bulk parameter $u_0$ and the bulk field $\Phi$
in terms of their renormalized counterparts
$u$ and $\Phi_{ren}$ are not affected by the presence of the
surface. Here $Z_u = 1 + {\cal O}(u)$ and 
$Z_{\Phi} = 1 + {\cal O}(u^2)$ are the corresponding
renormalization factors, and
$\mu$ is the inverse length scale which determines the 
renormalization-group flow. Since all surfaces we average
over are smooth on short distances, i.e., distances 
much smaller than the crossover length $\lambda$,
we expect that the surface operator $\Phi^{\perp}$ is renormalized 
by the same renormalization factor $Z_1$ which would occur 
for a {\em flat\/} surface with Dirichlet boundary 
conditions. Thus, 
\begin{equation} \label{sopren}
\Phi^{\perp} \, = \, (Z_{\Phi} Z_1)^{1/2} \, \Phi^{\perp}_{ren}
,
\end{equation}
with \cite{Die86}
\begin{equation} \label{Z1}
Z_1 \, = \, 1 \, + \, \frac{n+2}{3} \, \frac{u}{\varepsilon}
\, \, + \, {\cal O}(u^2) \, \, .
\end{equation}  
Using the above reparametrizations
the renormalized, i.e., pole-free, counterpart of 
$\overline{ {\cal G}_{\perp}}$
is given by
\begin{equation} \label{Fperpren}
\overline{ {\cal G}_{\perp} }_{,\,ren}
(\delta ; \omega, \lambda ; u , \mu) \, = \, 
Z_{\Phi}^{-1} Z_1^{-1/2} \,
\overline{{\cal G}_{\perp}}(\delta ; \omega , \lambda ;  u_0) \, \, .
\end{equation}
This perturbative result can be improved using standard
renormalization-group methods, by noting that
$\overline{{\cal G}_{\perp}}$ does not depend on $\mu$.
The asymptotic scaling behavior is governed by the infrared 
(long-distance) stable fixed point for which
\begin{equation} \label{ustar}
u = u^{\ast} \, = \, \frac{3 \, \varepsilon}{n + 8} \, \, + \,
{\cal O}(\varepsilon^2) \, \, ,
\end{equation} 
and $\overline{ {\cal G}_{\perp} }_{,\,ren}$
assumes the scaling form
\begin{equation} \label{fperp}
\overline{ {\cal G}_{\perp} }_{,\,ren}(\delta ; \omega, \lambda ; u, \mu)
\, \sim \, \delta^{-(d - 2 + \eta_{\perp})}
\, \, f_{\perp}( \delta / \lambda ; \omega / \lambda)
\end{equation} 
with the critical exponent $\eta_{\perp}$ for a flat surface. 
The scaling function 
$f_{\perp}$ is universal, but depends on the
particular way we have introduced the crossover length 
$\lambda$ in Eq.\,(\ref{crossover}). Since all surfaces we
average over are smooth on short distances,
$f_{\perp}(0; \omega / \lambda)$ should be a finite number
(in the following we suppress the dependence of $f_{\perp}$ on 
$\omega / \lambda$ for brevity).
In the other limit $\delta / \lambda \to \infty$, the scaling
function $f_{\perp}(\delta / \lambda)$ is expected to exhibit 
a power law which reflects the self-affine structure of the 
surface.

We have confirmed Eq.\,(\ref{fperp}) explicitly
to one-loop order according to Eq.\,(\ref{loop2}), 
using the expansion of $G(\underline{r} , \underline{r}')$
up to second order in $h({\bf x})$ in 
Eqs.\,(\ref{h0delta}) - (\ref{h2delta}),
and averaging using Eq.\,({\ref{crossover}).
Figure \ref{graph} illustrates this double expansion in 
graphical form \cite{rem}.
 
%%%%%%%%%%%%%%%%%%%%%%%% Fig.: graph.ps %%%%%%%%%%%%%%%%%%%%%%%%%

\unitlength1cm
\begin{figure}[t]
\begin{picture}(16,11.5)
\put(0.3,0.7){
\setlength{\epsfysize}{10cm}
\epsfbox{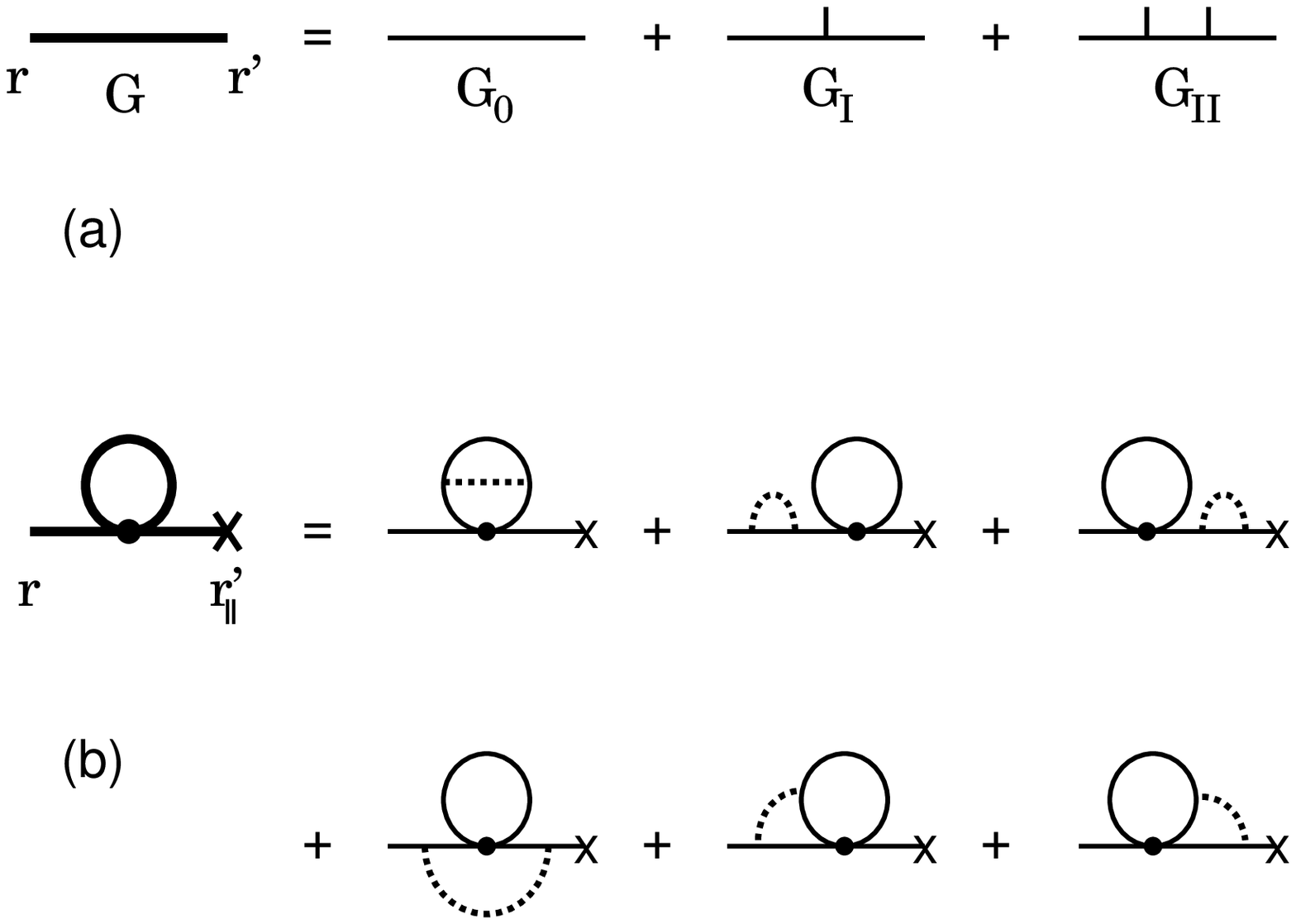}}
\end{picture}
\caption{(a) Representation of the full Gaussian propagator 
$G(\underline{r} ; \underline{r}')$ in Eq.\,(\ref{gauss2}) 
and its expansion up to second order in $h({\bf x})$ according to 
Eqs.\,(\ref{h0delta}) - (\ref{h2delta}). The number of ticks 
corresponds to the order in $h({\bf x})$. (b) The second order in 
$h({\bf x})$ contribution to the one-loop integral in 
Eq.\,(\ref{loop2}) decomposes into several parts. 
The dashed lines connecting the ticks indicate averaging
over different surface profiles, using Eq.\,(\ref{crossover}).
The cross corresponds to the surface operator
$\Phi^{\perp}$.}
\label{graph}
\end{figure}

%%%%%%%%%%%%%%%%%%%%%%%%%%%%%%%%%%%%%%%%%%%%%%%%%%%%%%%%%%%%%%%%%%%%

\smallskip

We indeed find that the $1 / \varepsilon$ poles generated 
by the surface operator $\Phi^{\perp}$ in Eq.\,(\ref{Fperp})
are removed by the 
renormalization factor $Z_1$ in Eq.\,(\ref{Z1}), which provides
a test of our calculation, and for the reasoning leading to 
Eq.\,(\ref{fperp}). This calculation gives also
the explicit form of the scaling function $f_{\perp}$ to 
first order in $\varepsilon$. We confirm, in particular, that
$f_{\perp}(0)$ is a finite number, and that the 
logarithmic contributions of $f_{\perp}(\delta / \lambda)$ 
for $\delta / \lambda \to \infty$ can be recast in the form 
of a power law, i.e.,
\begin{equation} \label{scaling}
f_{\perp}( \delta / \lambda ) \, \to \, 
\alpha \, + \,
\beta \, (\delta / \lambda)^{\psi} \, \, .
\end{equation}
Whereas both amplitudes $\alpha$ and $\beta$ depend on 
$\omega / \lambda$, the universal exponent $\psi$ 
is independent of $\omega / \lambda$ and given by
\begin{equation} \label{phiperp}
\psi \, = \, \frac{3}{2} \, \frac{n+2}{n+8} \, \varepsilon
\, - \, (2 - 2 \zeta) \quad + \, {\cal O}(\varepsilon^2) \, \, .
\end{equation}

Perpendicular correlations are obtained when $\underline{r}$ moves 
into the bulk, while $\underline{r}'$ remains close to the surface, 
i.e., $\delta \to \infty$ with $\delta'$ fixed (see Fig.\,\ref{fig1}). 
Equations (\ref{Fperp}) and (\ref{fperp}) - (\ref{phiperp}) then
imply that the correlations decay as
\begin{equation} \label{perp}
\overline{
\langle \Phi_i(\underline{r}) \Phi_i(\underline{r}') \rangle}
\, \sim \, \frac{1}{ \delta^{d-2+\eta_{\perp}}} \, + \,
\frac{a}{ \delta^{d-2+ \widetilde{\eta}_{\perp}}} \, \, ,
\end{equation}
where the first term corresponds to a flat surface with 
$\eta_{\perp} = 1 -  \frac{1}{2} \frac{n+2}{n+8} \varepsilon
\, + \, {\cal O}(\varepsilon^2)$ \cite{Bin83,Die86}. 
The second term describes the 
effect of self-affine roughness, with an amplitude $a$
depending on $\omega$, $\lambda$, and $\zeta$,
and the new universal exponent
\begin{equation} \label{etaperptilde}
\widetilde{\eta}_{\perp} \, = \, 
\eta_{\perp} - \psi \, = \,
(2 - 2 \zeta) \, + \, 
1 \, - \, 2 \, \frac{n+2}{n+8} \, \varepsilon
\, \, + \, {\cal O}(\varepsilon^2) \, \, .
\end{equation}
Similarly, when both points remain close to the surface, i.e., 
$\rho = |{\bf r}_{\parallel} - {\bf r}_{\parallel}'| \to \infty$
with $\delta$ and $\delta'$ fixed, the correlations decay as 
\begin{equation} \label{para}
\overline{
\langle \Phi_i(\underline{r}) \Phi_i(\underline{r}') \rangle}
\, \sim \, \frac{1}{ \rho^{d-2+\eta_{\parallel}}} \, + \,
\frac{a'}{ \rho^{d-2+ \widetilde{\eta}_{\parallel}}} \, \, .
\end{equation}
In this case the flat surface is governed by
$\eta_{\parallel} = 2 - \frac{n+2}{n+8} \, \varepsilon
+ {\cal O}(\varepsilon^2)$,
while self-affine roughness gives
\begin{equation} \label{etaparalleltilde}
\widetilde{\eta}_{\parallel}
\, = \, (2 - 2 \zeta) \, + \,
2 \, - \, 4 \, \frac{n+2}{n+8} \, \varepsilon \, \,
+ \, {\cal O}(\varepsilon^2) \, \, .
\end{equation}
The corrections due to roughness now decay with a slower
power as compared to the Gaussian case. Indeed, for a sufficiently
large roughness exponent $\zeta$, these corrections can even dominate 
the result for the flat surface. The borderline roughness exponent is
$\zeta_{\perp}^{\ast} = 1 - \frac{3}{4} \,
\frac{n+2}{n+8} \, \varepsilon \, + \, {\cal O}(\varepsilon^2)$
for perpendicular, and a different value of
$\zeta_{\parallel}^{\ast} \, = \, 1 \, - \, \frac{3}{2} \, 
\frac{n+2}{n+8} \, \varepsilon \, + \, {\cal O}(\varepsilon^2)$
for parallel correlations.
This is a surprising result from a naive point of view since,
due to $\zeta < 1$, on larger and larger length scales 
a self-affine rough surface looks more and more like a flat 
surface. Note that this effect becomes only visible beyond 
the Gaussian approximation, which corresponds to $\varepsilon = 0$.
By setting $\varepsilon = 1$ in the above expressions, 
one obtains the corresponding estimates for $d = 3$.
 
%%%%%%%%%%%%%%%%%%%%%%%%%%%%%%%%%%%%%%%%%%%%%%%%%%%%%%%%%%%%%%%%%%%%

\newpage

\section{Two dimensional XY model}
\label{sectionxy}

To compare the results of the previous Section with a different
interacting theory, we examine the correlations for a two dimensional 
XY model below the Kosterlitz-Thouless temperature \cite{KT78}. The 
order parameter in this system is the spin variable 
$s(\underline{r}) = e^{i \theta(\underline{r})}$, where
$\theta(\underline{r})$ is the angle (phase) the spin 
makes with some reference axis. Even though the phase fluctuations 
are described by a Gaussian model, nontrivial spin-spin correlations 
are obtained.
Below the Kosterlitz-Thouless temperature, the $d=2$ dimensional 
XY model is well described by the spin-wave Hamiltonian (neglecting 
vortices) \cite{Nel83}
\begin{equation} \label{sw}
\beta {\cal H}\{ \theta \} \, = \, 
\frac{1}{2} \, K \int d^2 r \, (\nabla \theta)^2 \, \, ,
\end{equation}
where $\underline{r} = (x,y)$.
Correlation functions decay as power laws in this system.
For instance, the two-point correlation function in the 
unbounded plane is given by \cite{Nel83,Car84}
\begin{equation} \label{sw2p}
G_b(\underline{r};\underline{r}') \, = \,
\langle 
e^{i \theta(\underline{r})} e^{- i \theta(\underline{r}')}
\rangle \, = \, 
\exp[{\cal G}_b(\underline{r}; \underline{r}') - 
\textstyle{\frac{1}{2}}{\cal G}_b(\underline{r}; \underline{r}) -
\textstyle{\frac{1}{2}}{\cal G}_b(\underline{r}'; \underline{r}')]
\end{equation}
with
\begin{equation} \label{swG}
{\cal G}_b(\underline{r}; \underline{r}') \, = \, \langle 
\theta(\underline{r}) \theta(\underline{r}') \rangle
\, = \, - \, \frac{1}{2 \pi K} \, \ln(r/a) \, \, ,
\end{equation}
where $r = |\underline{r} - \underline{r}'|$ and 
$a$ is some lattice cutoff. This implies
\begin{equation} \label{swexp}
G_b(\underline{r};\underline{r}') \, = \, 
\left( \frac{r}{a} \right)^{-\eta} \, \, ,
\end{equation}
where $\eta = 1/(2 \pi K)$. 

If the plane is bounded by a free surface (line) at $y = 0$, 
the correlation function $G(\underline{r};\underline{r}')$ 
in the half-space $y > 0$ is given by similar expressions 
as in Eq.\,(\ref{sw2p}), where now
${\cal G}(\underline{r}, \underline{r}')$
satisfies the Neumann boundary condition
at the surface \cite{Car84}. The final result
\begin{equation} \label{swhalf}
G(x,y;x',y') \, \sim \, 
\left[ \frac{[(x-x')^2 + (y-y')^2] [(x - x')^2 + (y + y')^2]}{4 y y'}
\right]^{- \eta/2}
\end{equation}
implies the surface critical exponents
$\eta_{\parallel} = 2 \eta$ and $\eta_{\perp} = \frac{3}{2} \eta$,
which fulfill the scaling relation 
$2 \eta_{\perp} - \eta_{\parallel} = \eta$ familiar from 
the surface critical behavior of $n$-vector models 
\cite{Bin83,Die86}.

In order to study whether the nontrivial roughness dependence
of correlations obtained in the previous Section 
is also present here, we now consider a deformed 
surface (line) with the same boundary conditions as above. Similar 
steps as outlined in Appendix \ref{appA} lead to the result for the
two-point correlation function 
\begin{equation} \label{swdeformed}
G(\underline{r};\underline{r}') \, = \, 
\exp[\Gamma(\underline{r}; \underline{r}') - 
\textstyle{\frac{1}{2}}\Gamma(\underline{r}; \underline{r}) -
\textstyle{\frac{1}{2}}\Gamma(\underline{r}'; \underline{r}')]
\end{equation}
with
\begin{equation} \label{swgamma}
\Gamma(\underline{r};\underline{r}') \, = \,
{\cal G}_b(\underline{r}; \underline{r}') - 
\int dx \int dx' \, 
\partial_n {\cal G}_b[\underline{r}; X(x)]
\, {\cal M}(x,x') \,
\partial_{n'} {\cal G}_b[\underline{r}'; X(x')] \, \, ,
\end{equation}
where $\partial_n$ denotes the normal derivative acting on $X$, 
and ${\cal M}(x,x')$ is the functional
inverse of $\partial_n \partial_{n'} {\cal G}_b[X(x), X(x')]$.
As in Sec.\,\ref{sectionfree},
we use the representation $X(x) = [x,h(x)]$ in terms of
the height profile $h(x)$, and expand 
$G(\underline{r};\underline{r}')$ up to second order in $h$.
In particular, for a self-affinely rough surface, we find, 
using Eq.\,(\ref{zeta}),
that the surface correlations fall off with the simple 
relative factor of $r^{-2(1-\zeta)}$ as compared to a 
flat surface (line) (compare Sec.\,\ref{sectionrough}).
We attribute this to the 
Gaussian nature of the fluctuations in the phase angle 
$\theta(\underline{r})$, which are retained in the asymptotics 
of correlations for $s(\underline{r})$.

%%%%%%%%%%%%%%%%%%%%%%%%%%%%%%%%%%%%%%%%%%%%%%%%%%%%%%%%%%%%%%%%%%%%

\section{Conclusion and outlook}
\label{sectiondiscussion}

We have developed a path-integral formulation for the study of
correlation functions in a system which is confined by a deformed
boundary. Our results are generic for any system with 
long-ranged correlations. Examples include systems with a broken
continuous symmetry, such as the XY model below the 
Kosterlitz-Thouless temperature, or a nematic liquid crystal,
where the correlations are generated by the corresponding 
massless Goldstone modes; or critical fluids or magnets 
described by the $n$-vector model at the bulk critical point,
which has been mostly considered in this work. The surface 
deformations can consist of specificly designed, regular 
patterns, or represent a self-affinely rough surface. Some conclusions
and possible extensions of this work are listed below.

(i) {\em Thermodynamic surface quantities\/}:
Thermodynamic quantities can be obtained from derivatives of the 
free energy with respect to magnetic fields. To discuss
surface behavior, we introduce distinct fields $h_b$ and $h_s$
in the bulk and close to the surface, respectively. 
Assuming that our 
underlying assumption of the validity of an expansion in 
$h({\bf x})$ holds, the results for the two-point correlation
function are consistent with the following form for the scaling 
of the leading singular part of the surface free energy 
per projected area, 
\begin{equation} \label{ansatz}
f_s^{(\text{sing})} \, = \, \xi^{-d+1\,}
\Big[ \, g_s(h_{b\,} \xi^{y_b}, h_{s\,} \xi^{y_s})
\, + \, \xi^{- 2 (1-\zeta)} \, g_r(h_{b\,} \xi^{y_b}, 
h_{s\,} \xi^{\widetilde{y}_s}) \Big] \, \, ,
\end{equation}
where $\xi \sim |T - T_c|^{- \nu}$
is the correlation length that diverges at the critical
point. The first term in square brackets corresponds to 
a flat surface, with $y_b$ and $y_s$ describing the relevance 
of bulk and surface fields, respectively \cite{Bin83,Die86}.
The second term gives the effect of surface roughness, with 
$\xi^{-2(1-\zeta)}$ reflecting the average increase in area.

By taking derivatives of Eq.\,(\ref{ansatz}), one can 
derive scaling relations between various surface critical exponents, 
in complete analogy to the case of a flat surface \cite{Bin83,Die86}.
In the following we focus on the contributions generated by the
surface roughness, which according to Eq.\,(\ref{ansatz})
appear {\em in addition\/} to the 
corresponding contributions for a flat surface. For example,
the singular part of the {\em surface magnetization\/}, 
$- \partial f_s^{(\text{sing})} / \partial h_s$,
can be written as $m_1 + \widetilde{m}_1$ so that
$\widetilde{\chi}_1 = \partial \widetilde{m}_1 / \partial h_b$
and 
$\widetilde{\chi}_{11} = \partial \widetilde{m}_1 / \partial h_s$
represent the contributions to the
{\em local susceptibility\/} and the 
{\em layer susceptibility\/} generated by the surface roughness,
respectively. Similarly, we suppose that the singular part of the 
two-point correlation function near the surface can be written as
$G(\underline{r}; \underline{r}')
+ \widetilde{G}(\underline{r}; \underline{r}')$, and
$\widetilde{G}(\underline{r}; \underline{r}')$
behaves for $h_b = h_s = 0$ as 
\begin{equation} \label{angle}
\widetilde{G}(\underline{r}; \underline{r}') \, \sim \,
\left\{ \begin{array} {l@{\quad\quad}l}
|\underline{r} - \underline{r}'|^{-(d-2+\widetilde{\eta}_{\parallel})} \, \,
\Gamma_{\parallel}(|\underline{r} - \underline{r}'| / \xi) \, \, , 
& \vartheta = 0 \, , \\[1mm]
|\underline{r} - \underline{r}'|^{-(d-2+\widetilde{\eta}_{\perp})} \, \,
\Gamma_{\perp}(|\underline{r} - \underline{r}'| / \xi, \vartheta) \, \, ,  
& \vartheta > 0 \, ,
\end{array} \right.
\end{equation}
where $\vartheta$ is the angle $\underline{r} - \underline{r}'$
makes with the surface, and $\Gamma_{\perp}$ vanishes for
$\vartheta \to 0$.
Equations (\ref{ansatz}) and (\ref{angle}) then imply the 
scaling relations between various critical exponents related 
to a rough surface shown in Table \ref{table1}.

Equations (\ref{etaperptilde}) and (\ref{etaparalleltilde}) for 
the $n$-vector model are consistent with the scaling relation for 
$\widetilde{\eta}_{\parallel}$ and 
$\widetilde{\eta}_{\perp}$ shown in Table \ref{table1}.
However, to regain the results in 
Eqs.\,(\ref{perp}) - (\ref{etaparalleltilde}),
we have to use a value of 
$\widetilde{y}_s = 1 + \frac{3 n}{2 (n+8)} 
\, \varepsilon \, + \, {\cal O}(\varepsilon^2)$
in Eq.\,(\ref{ansatz}), which is different from 
$y_s = 1 - \frac{3}{n+8} 
\, \varepsilon \, + \, {\cal O}(\varepsilon^2)$.
To motivate and justify this difference, we resort to
an analogy in which the rough surface is replaced with 
a collection of edges with a (possibly scale-dependent)
distribution of opening angles. Already for a single edge,
describing correlations requires a distinct
and angle-dependent value of $y_e$ for the magnetic field 
close to the edge \cite{Car83,HKSD99}. Similarly,
results obtained recently for correlations in the vicinity of a 
fractal surface with fractal dimension $d_f$ \cite{Dup98,Car99}
cannot be obtained using the value of $y_s$ for a flat surface
[with $\xi^{-d_f}$ replacing $\xi^{-d+1}$ in Eq.\,(\ref{ansatz})
and omitting the second term in square brackets].
Thus $\widetilde{y}_s$ can be regarded as inherently related 
to self-affine geometry. Interestingly, however, $\widetilde{y}_s$ 
itself does not depend on the roughness exponent $\zeta$, at least 
to order $\varepsilon$.

%%%%%%%%%%%%%%%%%%%%% Table %%%%%%%%%%%%%%%%%%%%%%%%%%%%%%%

{\center 
\begin{minipage}[t]{16cm}
\begin{table}[tbp]
\caption{Scaling relations between critical exponents relevant 
to a rough surface, as derived from Eqs.\,(\ref{ansatz}) and 
(\ref{angle}), in terms of the bulk critical exponents 
$\eta$, $\nu$, $y_b = \Delta / \nu$, 
and the roughness exponent $\zeta$.
For each exponent in the left column there is a corresponding
exponent for a flat surface [1,2] which would be 
denoted without tilde (compare with Table\,III in Ref.\,[2]).}
\label{table1}
\medskip
\begin{tabular}{lcc}
critical exponent & conditions & scaling relation \\ \hline
$\widetilde{\eta}_{\perp}$, $\widetilde{\eta}_{\parallel}$ 
\, \, Eqs.\,(\ref{perp}), (\ref{para}) & 
$\tau = h_b = h_s = 0$ & 
$2 \, \widetilde{\eta}_{\perp} - \widetilde{\eta}_{\parallel} =
\eta + 2 - 2 \zeta$ \\
$\widetilde{y}_s$ $\qquad \, \, \,$ Eq.\,(\ref{ansatz}) 
& $\tau \neq 0$, $h_s \neq 0$ &
$\widetilde{y}_s = 
\frac{1}{2} (d - \widetilde{\eta}_{\parallel} + 2 - 2 \zeta)$ \\
$\widetilde{\chi}_1 \sim |\tau|^{- \widetilde{\gamma}_1}$ & 
$\tau \neq 0$, $h_b = h_s = 0$ &
$\widetilde{\gamma}_1 = \nu (2 - \widetilde{\eta}_{\perp})$ \\
$\widetilde{\chi}_{11} \sim |\tau|^{- \widetilde{\gamma}_{11}}$ & 
$\tau \neq 0$, $h_b = h_s = 0$ &
$\widetilde{\gamma}_{11} = \nu (1 - \widetilde{\eta}_{\parallel})$ \\
$\widetilde{m}_1 \sim (- \tau)^{\widetilde{\beta}_1}$ & 
$\tau < 0$, $h_b = h_s = 0$ & 
$\widetilde{\beta}_1 = \frac{\nu}{2} 
(d - 2 + \widetilde{\eta}_{\parallel} + 2 - 2 \zeta)$ \\
$\widetilde{m}_1 \sim |h_b|^{1/\widetilde{\delta}_1}$ & 
$\tau = h_s = 0$, $h_b \neq 0$ & $\widetilde{\delta}_1 = 
\nu y_b / \widetilde{\beta}_1$ \\
$\widetilde{m}_1 \sim |h_s|^{1/\widetilde{\delta}_{11}}$ & 
$\tau = h_b = 0$, $h_s \neq 0$ & $\widetilde{\delta}_{11} = 
\nu \widetilde{y}_s / \widetilde{\beta}_1$
\end{tabular}
\end{table}
\end{minipage}\\[8mm]
}

%%%%%%%%%%%%%%%%%%%%%%%%%%%%%%%%%%%%%%%%%%%%%%%%%%%%%%%%

(ii) {\em Higher orders of the perturbation theory\/}: As the previous
remark already indicates, higher order results in $\varepsilon$ are
necessary in order to check the generality of our results for the 
$n$-vector model. For the contributions up to second order in 
$h({\bf x})$ (as considered here), we expect a systematic 
expansion in powers of $\varepsilon$, and one can calculate the 
${\cal O}(\varepsilon^2)$ contributions of, e.g., $f_{\perp}$ 
and $\psi$ in Eq.\,(\ref{scaling}). All the information needed 
about the self-affinely rough surface is contained in 
Eq.\,(\ref{crossover}). 
However, it is not clear how the perturbative calculation in 
$h({\bf x})$, for a self-affinely rough surface,
can be generalized to higher orders than the second. Such an 
attempt would require, in addition to Eq.\,(\ref{crossover}),
the knowledge of stochastic averages of three and more fields 
$h({\bf x})$, which can also introduce new length scales.  
Regarding these obstacles, it would be desirable to complement 
our results with a nonperturbative approach, e.g., for
the two-dimensional Ising model bounded by a self-affinely 
rough boundary. 

(iii) {\em Multiscaling\/}: For a random fractal boundary, it has been
shown \cite{Car99} that correlation functions exhibit multiscaling,
which means that the average (over fractal realizations of the boundary 
with given fractal dimension $d_f$) of their $n$th power does not scale 
in the same way as the $n$th power of their average. It would be 
interesting to see if similar behavior also applies to 
self-affine rough boundaries.

%%%%%%%%%%%%%%%%%%%%%%%%%%%%%%%%%%%%%%%%%%%%%%%%%%%%%%%%%%%%%%%%%%%%

\section*{Acknowledgements}

We thank J. Cardy, S. Dietrich, T. Emig, and R. Golestanian 
for helpful discussions and conversations.
This work was supported by the Deutsche Forschungs\-gemeinschaft
through Grant No. HA3030/1-2 (AH) and the National Science 
Foundation through Grants No. DMR-01-18213 and PHY99-07949 (MK).

%%%%%%%%%%%%%%%%%%%%%%%%%%%%%%%%%%%%%%%%%%%%%%%%%%%%%%%%%%%%%%%%%%%%

\begin{appendix}

\section{Path Integral formulation for correlation functions}
\label{appA}

In this Appendix we introduce a method to evaluate correlation 
functions of a fluctuating field subject to boundary conditions at 
surfaces of arbitrary shape. We consider a scalar field $\Phi$ 
described by the Gaussian action
\begin{equation} \label{action}
S\{ \Phi \} \, = \,   
\int d^d r \left[ \frac{1}{2} (\nabla \Phi)^2 \, + \,
\frac{\tau_0}{2} \, \Phi^2 \right] \, \, ,
\end{equation}
corresponding to Eq.\,(\ref{hamiltonian}) with $u_0 = 0$.
In order to study the behavior of correlation functions for
cases where more than one boundary surface is present, we
consider $N$ manifolds (objects) $\Omega_{\alpha}$ with
$\alpha = 1, \ldots, N$. Each point on the manifold 
$\Omega_{\alpha}$ is represented by a vector
$X_{\alpha}({\bf y}) = (X_{\alpha}^{\mu}({\bf y}); 
\mu = 1, \ldots, d)$.
Assuming the Dirichlet boundary condition $\Phi = 0$ on the manifolds,
a general correlation function with respect to the action (\ref{action})
can be written as
\begin{equation} \label{correlation}
\langle \, \bullet \, \rangle \, = \,
\frac{1}{Z_0} \int {\cal D} \Phi(\underline{r}) 
\prod_{\alpha=1}^{N} \prod_{X_{\alpha}} \delta[\Phi(X_{\alpha})]
\, \bullet \, e^{- S\{ \Phi \}} \, \, ,
\end{equation}
where
\begin{equation} \label{z0}
Z_0 \, = \, \int {\cal D} \Phi(\underline{r}) 
\prod_{\alpha=1}^{N} \prod_{X_{\alpha}} \delta[\Phi(X_{\alpha})]
\, \, e^{- S\{ \Phi \}} \, \, .
\end{equation}
Correlation functions of $\Phi$ can then be deduced from the
generating functional
\begin{equation} \label{generating}
Z\{J\} \, = \, \left\langle 
\exp\left[ \int d^d r J(\underline{r}) \Phi(\underline{r}) \right] 
\right\rangle \, \, , 
\end{equation}
which is normalized such that $Z\{0\} = 1$. 

Following Refs.~\cite{LK91,GK97}, we now express
for each manifold $\Omega_{\alpha}$ the boundary 
condition enforcing functional
$\prod_{X_{\alpha}} \delta[\Phi(X_{\alpha})]$ 
in terms of an auxiliary field $\Psi_{\alpha}(X_{\alpha})$ as
\begin{equation} \label{auxiliary}
\prod_{X_{\alpha}} \delta[\Phi(X_{\alpha})] \, \equiv \,
\int {\cal D} \Psi_{\alpha}(X_{\alpha}) \,
\exp\left[ i \int\limits_{\Omega_{\alpha}}
d X_{\alpha} \, \Psi_{\alpha}(X_{\alpha}) \Phi(X_{\alpha}) 
\right] \, \, .
\end{equation}
The Gaussian integration over $\Phi$ in 
Eqs.\,(\ref{correlation}), (\ref{z0}) can be 
performed, resulting in
\begin{equation} \label{resulting}
Z\{J\} \, = \, \mbox{const.} \, Z_b\{J\}
\int \prod_{\alpha = 1}^{N} 
{\cal D} \Psi_{\alpha}(X_{\alpha}) \,
e^{- \widetilde{S}_{\text{eff}}\{\Psi,J\}} \, \, ,
\end{equation}
where
\begin{equation} \label{zb}
Z_b\{J\} \, = \, \exp\left[  \frac{1}{2} \int d^d r \int d^d r' \, 
J(\underline{r}) \,
G_b(\underline{r}, \underline{r}') \,
J(\underline{r}') \right]
\end{equation}
with the bulk two-point correlation function
$G_b(\underline{r}, \underline{r}')$ corresponding to the 
action (\ref{action}). The effective action
$\widetilde{S}_{\text{eff}}$ is given by
\begin{eqnarray}
\widetilde{S}_{\text{eff}}\{\Psi,J\} \, & = & \,
\frac{1}{2} \, \sum_{\alpha \beta} 
\int\limits_{\Omega_{\alpha}} d X_{\alpha}
\int\limits_{\Omega_{\beta}} d X_{\beta} \, \,
\Psi_{\alpha}(X_{\alpha}) \,
G_b(X_{\alpha}, X_{\beta}) \,
\Psi_{\beta}(X_{\beta}) \label{effective}\\[2mm]
& & - \, i \, \sum_{\alpha} \int d^d r 
\int\limits_{\Omega_{\alpha}} d X_{\alpha} \,
J(\underline{r}) \,
G_b(\underline{r}, X_{\alpha}) \,
\Psi_{\alpha}(X_{\alpha}) \nonumber\, \, .
\end{eqnarray}
Note that evaluation of Eq.\,(\ref{resulting}) requires
functional integration over the curved manifolds $\Omega_{\alpha}$.
This is facilitated by expressing the
functional measure $\int {\cal D} \Psi_{\alpha}(X_{\alpha})$
in terms of the local coordinates 
${\bf y}$, which itself comprise a flat manifold (the local 
coordinate system). To this end we introduce the new fields 
$\psi_{\alpha}({\bf y}) \equiv \Psi_{\alpha}[X_{\alpha}({\bf y})]$.
However, this transformation requires some care regarding the 
integration measure $\int_{\Omega_{\alpha}} d X_{\alpha}$
in Eq.\,(\ref{effective}) as well as the functional measure 
$\int {\cal D} \Psi_{\alpha}(X_{\alpha})$ in 
Eq.\,(\ref{resulting}). The result is \cite{measure}
\begin{equation} \label{resultingnew}
\int \prod_{\alpha} 
{\cal D} \Psi_{\alpha}(X_{\alpha}) \,
e^{- \widetilde{S}_{\text{eff}}\{\Psi,J\}} \, = \,
\int \prod_{\alpha} {\cal D} \phi_{\alpha}({\bf y}) \,
e^{- S_{\text{eff}}\{\phi,J\}} \, \, ,
\end{equation}
where the field $\phi_{\alpha}({\bf y}) \equiv 
[g_{\alpha}({\bf y})]^{1/4} \psi_{\alpha}({\bf y})$ 
is given for each manifold $\Omega_{\alpha}$ in terms 
of the determinant $g_{\alpha}({\bf y})$ of its induced 
metric
\begin{equation} \label{metric}
g_{\alpha, ij}({\bf y}) \, = \, \sum_{\mu, \nu = 1}^{d} 
\frac{\partial X_{\alpha}^{\mu}}{\partial y_i}
\frac{\partial X_{\alpha}^{\nu}}{\partial y_j} \, \, .
\end{equation}
The new effective action $S_{\text{eff}}$ is given by
\begin{eqnarray}
S_{\text{eff}}\{\phi,J\} \, & = & \, \frac{1}{2} \, 
\sum_{\alpha \beta} 
\int d^D y \int d^D y' \,
\phi_{\alpha}({\bf y}) \, A_{\alpha \beta}({\bf y}, {\bf y}') \,
\phi_{\beta}({\bf y}') \label{effectivenew} \\[2mm]
& & - \, i \, \sum_{\alpha} \int d^d r \int d^D y \, 
J(\underline{r}) \, w_{\alpha}(\underline{r}, {\bf y}) \,
\phi_{\alpha}({\bf y}) \nonumber
\end{eqnarray}
with the kernels
\begin{mathletters} \label{MW}
\begin{eqnarray}
A_{\alpha \beta}({\bf y}, {\bf y}') \, & = & \,
[g_{\alpha}({\bf y})]^{1/4} \,
G_b[X_{\alpha}({\bf y}), X_{\beta}({\bf y}')] \,
[g_{\beta}({\bf y}')]^{1/4} \label{M} \, \, , \\[2mm]
w_{\alpha}(\underline{r}, {\bf y}) \, & = & \,
G_b[\underline{r}, X_{\alpha}({\bf y})] \,
[g_{\alpha}({\bf y})]^{1/4} \, \, . \label{W}
\end{eqnarray}
\end{mathletters}
The functional measure 
$\int {\cal D} \phi_{\alpha}({\bf y})$
on the right hand side of Eq.\,(\ref{resultingnew}) is the 
one conventionally used on a flat manifold. 
The corresponding Gaussian integrations can thus 
be performed, resulting in
\begin{equation} \label{result}
Z\{J\} \, = \, Z_b\{J\} \,
\exp\left[ - \, \frac{1}{2}       
\, \int d^d r \int d^d r'
J(\underline{r}) \,
K(\underline{r}, \underline{r}') \,
J(\underline{r}') \right]
\end{equation}
with the kernel
\begin{equation} \label{Ka}
K(\underline{r}, \underline{r}') \, = \, 
\sum_{\alpha \beta} \int d^D y \int d^D y' \,
w_{\alpha}(\underline{r}, {\bf y}) \,
A^{-1}_{\alpha \beta}({\bf y}, {\bf y}') \,
w_{\beta}(\underline{r}', {\bf y}') \, \, .
\end{equation}
Using 
$A^{-1}_{\alpha \beta}({\bf y}, {\bf y}') =
[g_{\alpha}({\bf y})]^{-1/4}
M_{\alpha \beta}({\bf y}, {\bf y}') \,
[g_{\beta}({\bf y}')]^{- 1/4}$,
where $M_{\alpha \beta}({\bf y}, {\bf y}')$ is the functional
inverse of $G_b[X_{\alpha}({\bf y}), X_{\beta}({\bf y}')]$
(with respect to both ${\bf y}$, ${\bf y}'$ and the indices 
$\alpha$, $\beta$), one finds that the factors of 
$[g_{\alpha}({\bf y})]^{1/4}$ in Eq.\,(\ref{Ka}) cancel.
From Eqs.\,(\ref{MW}) - (\ref{Ka})
one can thus read off the final result for the two-point 
correlation function,
\begin{equation} \label{two} 
G(\underline{r}, \underline{r}') \, = \,
G_b(\underline{r}, \underline{r}') \, - \,
\sum_{\alpha, \beta = 1}^{N} \int d^D y \int d^D y' \, 
G_b[\underline{r}, X_{\alpha}({\bf y})] \,
M_{\alpha \beta}({\bf y}, {\bf y}') \,
G_b[\underline{r}', X_{\beta}({\bf y}')] \, \, .
\end{equation}
Choosing $N = 1$, corresponding to only one manifold,
gives Eq.\,(\ref{gauss2}).

%%%%%%%%%%%%%%%%%%%%%%%%%%%%%%%%%%%%%%%%%%%%%%%%%%%%%%%%%%%%%%%

\section{Short distance expansion of the stress tensor} 
\label{appB}

In this Appendix we consider the expansion of the two-point 
correlation function for a general massless field theory described by a 
Hamiltonian ${\cal H}\{\Phi\}$, to first order in the deformations
of the height profile of a bounding surface. To this end, we 
introduce a new type of short-distance expansion of the stress 
tensor near a surface with the following scale-invariant boundary 
conditions: 
(a) the Dirichlet boundary condition $\Phi = 0$ corresponding to 
the ordinary surface universality class, and 
(b) the boundary condition $\Phi = \infty$ describing critical 
adsorption, corresponding to the extraordinary universality class.

A deformed surface $S$ given by the height profile $h({\bf x})$ 
(see Fig.\,\ref{fig1}) can be obtained from the flat surface $S_0$ 
with $h({\bf x}) = 0$ by means of a coordinate transformation, 
which maps the space $({\bf x}, z)$ on the space 
$(\widehat{\bf x}, \widehat{z})$. 
We define this transformation by
\begin{equation} \label{trafo}
\widehat{{\bf x}} \, = \, {\bf x} \, \, , \quad \, 
\widehat{z} \, = \, z \, + \, h({\bf x}) \, \Theta(z) \, \, ,
\end{equation}
where $\Theta(z)$ is an arbitrary differentiable function with
$\Theta(z) = 1$ for $z \leq z_0$ with some $z_0 > 0$, and which 
vanishes for $z \to \infty$. We denote the Hamiltonian with the 
flat surface $S_0$ by ${\cal H}$ and the Hamiltonian with a 
deformed surface $S$ by $\widehat{\cal H}$. According to the 
definition of the stress tensor $T_{ik}$ \cite{Brown80,Car87,ER95}
the change of ${\cal H}$ generated by the coordinate transformation 
(\ref{trafo}) can be written as
\begin{equation} \label{change}
\widehat{\cal H} - {\cal H} \, = \, - \,
\int_{\text{HS}} d^d r \, 
\sum_{k = 1}^{d} 
\left[\frac{\partial}{\partial r_k}
\Big( h({\bf x}) \, \Theta(z) \Big) \right] 
\, T_{zk}(\underline{r}) \, \, + \, {\cal O}( h^2 ) \, \, ,
\end{equation}
where ${\text{HS}}$ denotes the half-space $\underline{r} = ({\bf x}, z)$
with $z \geq 0$. Using the property $\sum_k \partial_k T_{ik} = 0$ 
and the divergence theorem, one obtains
\begin{equation} \label{div}
\widehat{\cal H} \, = \, {\cal H} \, + \, 
\int_{{\Bbb R}^{D}} d^D x \, h({\bf x}) \, T_{zz}({\bf x}, z=0)
\, \, + \, {\cal O}( h^2 ) \, \, .
\end{equation}
The contribution to first order in $h$ is located 
{\em at} the (flat) surface and does not depend on the specific 
choice of $\Theta(z)$. The higher order contributions 
${\cal O}( h^2 )$ cannot be transformed in this way, and 
will not be addressed in the following.   
$T_{zz}({\bf x},0) = \lim_{\delta \to 0} T_{zz}({\bf x}, \delta)$
represents a surface operator, which does not, however, need to be 
renormalized at the surface, so that its scaling dimension equals 
its canonical inverse length dimension of $d$ \cite{Car90,ES94}.

In the following we consider the cumulant  
$\langle \Phi(\underline{r}) \Phi(\underline{r}') \rangle^C$
of the two-point correlation function in the system 
described by ${\cal H}\{\Phi\}$ above the deformed surface $S$. 
Using Eq.\,(\ref{div}) one finds
\begin{equation} \label{cum}
\langle \Phi(\underline{r}) \Phi(\underline{r}') \rangle^C \, = \,
\langle \Phi(\underline{r}) \Phi(\underline{r}') \rangle_0^C \,
- \, \int d^D x \, h({\bf x}) \, \langle T_{zz}({\bf x}, 0) \,
\Phi(\underline{r}) \Phi(\underline{r}') \rangle_0^C \, \, + \, {\cal O}(h^2)
,
\end{equation}
where $\langle \, \, \, \, \rangle_0^C$ denotes the cumulant 
within the half-space $\text{HS}$ bounded by the flat surface 
$S_0$. We now consider the limit
$\rho = | {\bf r}_{\parallel} - {\bf r}_{\parallel}' | \to \infty$
(see Fig.\,\ref{fig1}), so that we can use the short-distance expansion 
(SDE) of the order parameter $\Phi(\underline{r})$ near the surface. 
For the first term 
$\langle \Phi(\underline{r}) \Phi(\underline{r}') \rangle_0^C$ 
in Eq.\,(\ref{cum}), the SDE is well-known:
(a) for the Dirichlet boundary condition $\Phi = 0$, the SDE is 
given by \cite{Bin83,Die86}
\begin{mathletters} \label{sd}
\begin{equation} \label{sdo}
\Phi({\bf r}_{\parallel}, z) \, = \, a \, 
z^{(\eta_{\parallel} - \eta) / 2}
\, \frac{\partial}{\partial z} \Phi({\bf r}_{\parallel}, z) 
\Big|_{z = 0} \, + \, \ldots \, \, , 
\end{equation}
where $\partial_z \Phi({\bf r}_{\parallel}, z = 0)$ is  
a surface operator, $\eta_{\parallel}$ is a surface critical
exponent, and $a$ is a nonuniversal amplitude; 
(b) for the boundary condition $\Phi = \infty$,
the SDE has the form \cite{Car90,ES94}
\begin{equation} \label{sde}
\frac{\Phi({\bf r}_{\parallel}, z)}
{\langle \Phi({\bf r}_{\parallel},z) \rangle_0}
\, = \, I \, + \, b_T \, z^{d} \,  
T_{zz}({\bf r}_{\parallel}, z = 0)  \, + \, \ldots \, \, ,
\end{equation}
\end{mathletters}
where $\langle \Phi ({\bf r}_{\parallel},z) \rangle_0$
is taken at the critical point of the field theory, and 
$I$ is the identity operator. The amplitude $b_T$ is universal.
Equation (\ref{sd}) in conjunction with the scaling behavior 
$\langle \Phi(\underline{r}) \Phi(\underline{r}') \rangle_0^C \sim
\rho^{-(d-2+\eta)} \, f(\frac{z}{\rho}, \frac{z'}{\rho})$
\cite{Bin83,Die86} gives the result for a flat surface
\begin{equation} \label{def_eta}
\langle \Phi(\underline{r}) \Phi(\underline{r}') \rangle_0^C \, \sim \,
(z z')^{(\eta_{\parallel} - \eta) / 2}
\rho^{- (d - 2 + \eta_{\parallel})} \, \, , \quad \rho \to \infty \, \, .
\end{equation}
For boundary condition (b), one has $\eta_{\parallel} = d + 2$ 
\cite{Car90,ES94}, and the property $\langle T_{zz} \rangle = 0$ 
has been used.

For the second term 
$\int d^D x \, h({\bf x}) \, \langle T_{zz}({\bf x}, 0) \,
\Phi(\underline{r}) \Phi(\underline{r}') \rangle_0^C$ on the 
right hand side (rhs) of
Eq.\,(\ref{cum}), the above procedure cannot be applied
directly because the integration of $T_{zz}({\bf x}, 0)$
separates the points $\underline{r}$ and $\underline{r}'$ from the 
surface. To proceed, it is illustrative to consider first
the case of a {\em constant\/} height field $h({\bf x}) = h_0$.
In this case, the integration of $T_{zz}({\bf x}, 0)$ simply
amounts to a surface shift in the form \cite{DDE83}
\begin{equation} \label{shift}
h_0 \int d^D x \, \langle T_{zz}({\bf x}, 0) \,
\Phi(\underline{r}) \Phi(\underline{r}') \rangle_0^C \, = \, 
h_0 \left( 
\frac{\partial}{\partial z} + \frac{\partial}{\partial z'}
\right) \, 
\langle \Phi(\underline{r}) \Phi(\underline{r}') \rangle_0^C \, \, .
\end{equation}
Consider for illustration the case for which only 
$\underline{r} = ({\bf r}_{\parallel}, z)$ is close to 
the surface, i.e., $z \ll z'$. Since 
$\langle \Phi(\underline{r}) \Phi(\underline{r}') \rangle_0^C$
for small $z$ behaves like a power in $z$, 
the $z$ derivative on the rhs of 
Eq.\,(\ref{shift}) is larger than 
the $z'$ derivative by an amount of order $z'/z$. 
Now we recall that for the 
boundary conditions (a) and (b), correlations near 
the surface are {\em suppressed\/}, so that one can expect that
on the left hand side of Eq.\,(\ref{shift}) actually only a small 
integration region around ${\bf r}_{\parallel}$ contributes
to the $z$ derivative on the rhs. This suggests the 
operator product expansion
\begin{equation} \label{new}
T_{zz}({\bf x}, 0) \, \Phi( {\bf r}_{\parallel}, z) \, = \,
\Delta({\bf x} - {\bf r}_{\parallel}, z) \, 
\frac{\partial}{\partial z} \Phi( {\bf r}_{\parallel}, z) \, + \, 
\ldots
\end{equation}
for $({\bf x}, 0)$ close to $\underline{r} = ({\bf r}_{\parallel}, z)$,
where $\Delta({\bf x}, z)$ is a representation of the delta function 
$\delta^D({\bf x})$ in $D$ dimensions, i.e.,
\begin{equation} \label{delta}
\int d^D x \, \Delta({\bf x}, z) = 1 \, \, \, ,
\quad \lim_{z \to 0} \, \Delta({\bf x}, z)
\, = \, \delta^D({\bf x}) \, \, .
\end{equation}
Note that $\partial_z \Phi( {\bf r}_{\parallel}, z)$ on the 
rhs of Eq.\,(\ref{new}) is {\em not\/} a surface operator, since 
the $z$ derivative is taken at a distance $z > 0$ from the surface.  
The validity of Eq.\,(\ref{new}) can be verified for various
cases. For two-dimensional systems at criticality bounded by a line 
with the boundary condition (a) or (b), it follows from the local 
form of the conformal Ward Identity \cite{Car84}.   
For the Gaussian model with the boundary condition (a),
it can easily be verified for any dimension $d$. 
For a $\Phi^4$ model at criticality with boundary condition (b), 
Eq.\,(\ref{new}) is consistent with the form of 
$\langle T_{zz}({\bf x}, 0) \, \Phi( {\bf r}_{\parallel}, z) \rangle_0$ 
known from conformal invariance arguments for 
any $d$ with $2 \le d \le 4$ \cite{Car90,ES94}.
For this system we have checked Eq.\,(\ref{new}) also for the 
correlation function 
$\langle \varphi(\underline{r}) \varphi(\underline{r}') \rangle_0$ with
$\varphi(\underline{r}) = 
\Phi(\underline{r}) - \langle \Phi(\underline{r}) \rangle$
\cite{ES94} to first (one loop) order in the $\Phi^4$ interaction.

Let us go back to the second term
$\int d^D x \, h({\bf x}) \, \langle T_{zz}({\bf x}, 0) \,
\Phi(\underline{r}) \Phi(\underline{r}') \rangle_0^C$ on the rhs of
Eq.\,(\ref{cum}) with $z$ and $z'$ fixed.
In order to obtain its leading behavior for 
$\rho \to \infty$, the ${\bf x}$ integration 
can be divided in two regions. Within one region, 
${\bf x}$ is far away from both 
${\bf r}_{\parallel}$ and ${\bf r}_{\parallel}'$.
Hence Eq.\,(\ref{sd}) can be applied to both points
${\bf r}_{\parallel}$ and ${\bf r}_{\parallel}'$.
Within the complement region, ${\bf x}$ is either close 
to ${\bf r}_{\parallel}$ or to ${\bf r}_{\parallel}'$
so that Eq.\,(\ref{new}) can be used. Due to the 
$z$ derivative in Eq.\,(\ref{new}) in conjunction with the
scaling behavior quoted below Eq.\,(\ref{sd}), the 
contribution arising from the second integration
region is by a factor $\rho / z$ or $\rho / z'$ larger 
than the contribution from the first integration region.
Using Eq.\,(\ref{def_eta}), we conclude that the leading 
contribution for $\rho \to \infty$ of the second term on 
the rhs of Eq.\,(\ref{cum}) is given by
$[A(\underline{r}) + A(\underline{r}')]
\langle \Phi(\underline{r}) \Phi(\underline{r}') \rangle_0^C$
with the amplitude $A(\underline{r})$ in Eq.\,(\ref{A1}).
Thus we obtain the leading scaling behavior of
$\langle \Phi(\underline{r}) \Phi(\underline{r}') \rangle^C$
for $\rho \to \infty$ quoted in Eq.\,(\ref{leading}).

%%%%%%%%%%%%%%%%%%%%%%%%%%%%%%%%%%%%%%%%%%%%%%%%%%%%%%%%%%%%%%%

\section{Structure of the loop expansion} 
\label{appC}

We consider the diagrams on the right hand side
of Fig.\,\ref{graph}\,(b). 
According to Eq.\,(\ref{loop2}), in the $({\bf p}, \delta)$
representation, the distances $\delta_0$ of the $\Phi^4$ vertices 
[dots in Fig.\,\ref{graph}\,(b)] from the surface have to be integrated 
using $\int_0^{\infty} d \delta_0$. To one loop order, only the three
diagrams in the first line of Fig.\,\ref{graph}\,(b) exhibit 
short-distance singularities at $\delta_0 = 0$. These diagrams 
consist of the following components:
\begin{equation} \label{leftline}
\Leftline \, = \, g_0(p; \delta, \delta_0) \quad
\text{[see Eq.\,(\ref{half2})]} \, \, ;
\end{equation}
\begin{equation} \label{rightline}
\Rightline \, = \, e^{-p \delta_0} \, \, ;
\end{equation}
\begin{equation} \label{leftdashline}
\Dashleftline \, = \, g_2(p; \delta, \delta_0) \quad
\text{[see Eq.\,(\ref{G2})]} \, \, ;
\end{equation}
\begin{equation} \label{rightdashline}
\Dashrightline \, = \,
- \, p \, {\cal K}(p, 0) \, e^{-p \delta_0} \, + \,
p \, {\cal K}(p, \delta_0) \, - \,
\Omega \, e^{-p \delta_0}
\end{equation}
with the constant
$\Omega = \frac{\partial}{\partial \delta} \,
{\cal K}(p, \delta)_{\delta = 0\,}$;
\begin{equation} \label{loop}
\Loop \, = \, {\cal A} \, \delta_0^{1-D}
\end{equation}
with the constant
\begin{equation} \label{A}
{\cal A} \, = \, - \int \frac{d^D \alpha}{(2 \pi)^D} \, 
\frac{1}{2 \alpha} \, e^{- 2 \alpha} \, \, ;
\end{equation}
\begin{eqnarray} \label{dashloop} 
\Dashloop \, & = & \int \frac{d^D p}{(2 \pi)^D} \,  
\left[{\cal K}(p, 0) \, e^{-2 p \delta_0} \, - \, 
2 \, {\cal K}(p, \delta_0) \, e^{-p \delta_0} \right] \\[2mm]
& = & \, \Omega \, {\cal B} \, \delta_0^{1-D} \, + \, 
F_1(\delta_0) \, \, , \nonumber
\end{eqnarray}
where the function $F_1(\delta_0)$ is regular for $\delta_0 \to 0$.
The constant ${\cal B}$ is given by
\begin{equation} \label{B}
{\cal B} \, = \, D^{-1} \int \frac{d^D \alpha}{(2 \pi)^D} \, 
\left[\widetilde{U}_0(\alpha) \, e^{-2 \alpha} \, - \, 
2 \, \widetilde{U}(\alpha) \, e^{-\alpha} \right] \, \, ,
\end{equation}
where $\widetilde{U}_0(p_{} \delta) = U(p,0) / \delta$ and 
$\widetilde{U}(p_{} \delta) = U(p,\delta) / \delta$, with
$U(p,\delta)$ from Eq.\,(\ref{Up}).
Note that in Eq.\,(\ref{dashloop}) the terms in square brackets
in Eq.\,(\ref{G2}), which correspond to the first line in
Eq.\,(\ref{h2delta}), do not contribute.

Writing
${\cal A} = {\cal A}_0 + \varepsilon {\cal A}_1 + {\cal O}(\varepsilon^2)$ and
${\cal B} = {\cal B}_0 + \varepsilon {\cal B}_1 + {\cal O}(\varepsilon^2)$
with $\varepsilon = 4 - d$,
one finds that ${\cal A}_0 = {\cal B}_0$.
This nontrivial fact is the  
reason why the $1/\varepsilon$ poles due to the short-distance 
singularities of the first 
and the third diagram in the first line of Fig.\,\ref{graph}\,(b)  
cancel each other. The second diagram can be written as
${\cal A} \, [1/\varepsilon - C_E - \ln(p)]
\, \Dashrightline \, + F_2(p,\delta)$, with Euler's constant $C_E$
and a pole-free function $F_2(p,\delta)$.
The $1/\varepsilon$ pole in this expression 
is then removed by the factor
$Z_1^{-1/2}$, with $Z_1$ from Eq.\,(\ref{Z1}), that 
multiplies the zero loop contribution $\Dashrightline\,$ of 
the correlation function. The remaining, regular contributions,
including those from the diagrams in the second line of 
Fig.\,\ref{graph}\,(b), contain additional logarithmic terms 
in $\delta$ which are not present if the surface was flat. 
One can then identify these logarithmic contributions, and 
show that they can be recast in the power law according to 
Eqs.\,(\ref{fperp}) - (\ref{phiperp}). 
It should be noted, however, that here this exponentiation is 
not entirely based on an RG argument, but relies on the plausible 
assumption that the self-affine structure of the surface should 
result in pure power laws (without logarithmic corrections) for 
the decay of correlation functions. 

An analogous calculation leads to the quoted results for 
lateral correlations. Since in this case both points are 
located near the surface, only four of the six diagrams
in Fig.\,\ref{graph}\,(b) are different from each other.

\end{appendix}

\newpage

%%%%%%%%%%%%%%%%%%%%%%%%%%%%%%%%%%%%%%%%%%%%%%%%%%%%%%%%%%%%%%%


\begin{references}
\bibitem{rem}For self-affine rough surfaces, we can neglect 
correlations which contain $\nabla h$. This implies that we can use 
$\Phi^{\perp}({\bf r}_{\parallel}') =  
\partial \Phi(\underline{r}') / \partial \delta' |_{\delta'=0}$
for the surface operator [see Eq.\,(\ref{sop})].
\end{references}
\end{document}